\documentclass[a4paper,12pt]{article}
\pdfoutput=1 

\usepackage{jcappub} 
\usepackage{xcolor}
\usepackage{slashed}
\usepackage{subfigure}
\usepackage[T1]{fontenc}
\usepackage{ulem}

\def\s{\sigma}
\def\ra{\rangle}
\def\la{\langle}

\def\M{\mathcal{M}}

\def\c{\chi}
\def\s{\sigma}
\def\ap{\text{\it A$^\prime$}}
\def\G{\Gamma}
\def\to{\rightarrow}

\title{MeV Dark Matter Complementarity and the Dark Photon Portal}

\author{Ma\'{\i}ra Dutra$^1$,}
\emailAdd{maira.dutra@th.u-psud.fr}
\author{Manfred Lindner$^2$,}
\emailAdd{manfred.lindner@mpi-hd.mpg.de}
\author{Stefano Profumo$^3$,}
\emailAdd{profumo@ucsc.edu}
\author{Farinaldo S.\ Queiroz$^{4}$,}
\emailAdd{farinaldo.queiroz@iip.ufrn.br}
\author{Werner Rodejohann$^2$,}
\emailAdd{werner.rodejohann@mpi-hd.mpg.de}
\author{Clarissa Siqueira$^{2,5}$}
\emailAdd{clarissa@mpi-hd.mpg.de}

\affiliation{$^1$ 
Laboratoire de Physique Th\'eorique, CNRS -- UMR 8627, \\
Universit\'e de Paris-Saclay 11, F-91405 Orsay Cedex, France}
\affiliation{$^2$Max-Planck-Institut f\"ur Kernphysik, Postfach 103980, 69029 Heidelberg, Germany}
\affiliation{$^3$Department of Physics and Santa Cruz Institute for Particle Physics, University of California, Santa Cruz, CA 95064, USA}
\affiliation{$^4$International Institute of Physics, Federal University of Rio Grande do Norte, Campus Universit\'ario, Lagoa Nova, Natal-RN 59078-970, Brazil}
\affiliation{$^5$Departamento  de  F\'{\i}sica,  Universidade  Federal  da  Para\'{\i}ba, Caixa  Postal  5008,  58051-970,  Jo\~ao  Pessoa,  PB,  Brasil}

\abstract{
We discuss the phenomenology of an MeV-scale Dirac fermion coupled to the Standard Model through a dark photon with kinetic mixing with the electromagnetic field. We compute the dark matter relic density and explore the interplay of direct detection and accelerator searches for dark photons. We show that precise measurements of the temperature and polarization power spectra of the Cosmic Microwave Background Radiation lead to stringent constraints, leaving a small window for the thermal production of this MeV dark matter candidate. The  forthcoming MeV gamma-ray telescope e-ASTROGAM will offer important and complementary opportunities to discover dark matter particles with masses below $\sim 10$~MeV. Lastly, we discuss how a late-time inflation episode and freeze-in production could conspire to yield the correct relic density while being consistent with existing and future constraints. 
}

\begin{document}
\maketitle
\flushbottom

\section{Introduction}

The existence of dark matter (DM) has been established, via its gravitational effects, through a variety of
observations at different scales -- from galaxies to the largest structures in the Universe. The dark matter accounts for 27\% of the total energy density of the Universe, and for about 85\% of its matter density \cite{Ade:2015xua}. However, at present the fundamental nature of the DM particle remains a mystery, providing one of the most important open problems in particle and astroparticle physics today. The requirements of strong and electromagnetic charge neutrality, and of the dark matter being non-relativistic at the time of decoupling from the thermal bath in the early Universe, rule out any of the Standard Model known particles as dark matter candidates. The dark matter is thus quite likely a new elementary constituent. As such, it is not implausible to assume the dark matter is charged under possibly additional, ``dark'' gauge interactions.\\

In the Standard Model, electromagnetic interactions are described by a $U(1)$ gauge interaction, quantum electrodynamics (QED),  whose massless force carrier is the photon. In QED the photon couples to particles proportionally to their electric charges and features only vector interactions. One of the few renormalizable ``portals'' between the dark matter and the (visible) Standard Model is via the kinetic mixing of the electromagnetic field strength and the field strength of a new (dark) $U(1)$ gauge interaction, whose force carrier is a new particle \cite{FAYET1990743}, which we hereafter indicate as the {\it dark photon}. In principle the dark photon can be massless, with an unbroken dark $U(1)$ and a possible milli-charged DM \cite{Ackerman:mha,Knapen:2017xzo}. Here we will consider instead the possibility that the dark $U(1)$ is broken, leading to a massive dark photon. Searches for dark photons have been carried out at a multitude of laboratories throughout the world, utilizing data ranging from collisions at the Large Hadron Collider to pion decays \cite{Fayet:1980rr,Fayet:1980ss,Fayet:2006xd,An:2014twa,Lees:2014xha,Batley:2015lha,Aguilar-Arevalo:2016zop,Fayet:2016nyc,Aaij:2017rft,Ablikim:2017aab,Lees:2017lec,Choi:2017kzp}. \\

In our work, we assess the feasibility of having the dark photon as a portal to dark matter, assuming that the dark matter particle is an MeV-scale Dirac fermion charged under the new dark $U(1)$ gauge interaction. Dark matter at the MeV scale is an interesting possibility, offering a rich phenomenology \cite{Knapen:2017xzo}. The topic has witnessed increasing interest in light of null results in the search for WIMPs \cite{Arcadi:2017kky} and of the many upcoming  experimental probes for MeV-scale dark matter \cite{Kile:2009nn,Izaguirre:2013uxa,Izaguirre:2015pva,Hochberg:2017wce,Mei:2017etc,An:2017ojc}. Several MeV dark matter studies have been conducted in the literature, 
e.g.\ in the context of a light dark Higgs \cite{Darme:2017glc}, effective operators \cite{Boudaud:2016mos,Choudhury:2017osc,Bertuzzo:2017lwt}, radiative neutrino masses \cite{Arhrib:2015dez}, sterile neutrinos \cite{Huang:2013zga}, neutrino detectors such as Super-Kamiokande \cite{PalomaresRuiz:2007eu}, dark radiation \cite{Ho:2012ug}, interplay with gamma-rays \cite{Boddy:2015efa,Gonzalez-Morales:2017jkx}, in connection to MeV anomalies at colliders \cite{Chen:2016tdz}, supernova physics \cite{Fayet:2006sa,PhysRevD.73.103518,Dreiner:2013mua}, small scale structure \cite{Hooper:2007tu}, keV line emission \cite{Boehm:2003ha,Frere:2006hp,Fayet:2006sp,Fayet:2007ua}, low energy colliders \cite{Borodatchenkova:2005ct}, Big Bang Nucleosynthesis \cite{Serpico:2004nm} and coherent neutrino-nucleus scattering \cite{Ge:2017mcq}. 
  Despite the large number of existing studies in the context of MeV dark matter, our work is novel at a variety of levels:
\begin{itemize}
\item[(i)] We discuss complementarity of dark matter searches, focusing on accelerator and/or collider searches as well as direct and indirect dark matter searches;
\item[(ii)] We investigate MeV dark matter complementarity in the context of the dark photon portal;
\item[(iii)] We study different production mechanisms beyond thermal freeze-out, namely inflaton decay and freeze-in.
\end{itemize}

The paper is structured as follows: In Section \ref{sec:model}, we introduce our dark photon model; we discuss in Section \ref{sec:dm} the dark matter relic density; in Sections \ref{sec:cmb}-\ref{sec:gam}-\ref{sec:DD} we discuss the CMB and gamma-ray  constraints and prospects, as well as direct detection bounds; in Section \ref{sec:alt} present our results in the context of thermal production, late-time inflation and freeze-in production mechanisms, and we conclude in Section \ref{sec:concl}.


\section{\label{sec:model}The dark photon model}

The dark photon model has been originally proposed in \cite{Pospelov:2007mp,Pospelov:2008zw}. The model contains  a massive vector boson mixing with the QED photon via a kinetic mixing term of the form $\varepsilon F^{\mu \nu} F_{\mu \nu}^\prime$, with $\varepsilon$ a dimensionless parameter. In the regime in which the dark photon is much lighter than the $Z$ boson, the dark photon inherits the properties of the QED photon, i.e.\ it interacts with fermions proportionally to their electric charges, albeit with couplings suppressed by $\varepsilon$.  In this work, such dark photon is the mediator between the dark matter particle, assumed to be a Dirac fermion, and the SM fermions. After kinetic mixing diagonalization \cite{Alexander:2016aln}, the Lagrangian reads
\begin{equation}\begin{split}
\mathcal{L} \supset & -\frac{1}{4}F_{\mu \nu}F^{\mu \nu} -\frac{1}{4}F^\prime_{\mu \nu}F^{\prime\mu \nu} + \frac{1}{2} M_{A^\prime}^2 A'^2 \\
& + \sum_i \overline{f}_i(-e q_{f_i} \slashed{A} - \varepsilon e q_{f_i} \slashed{A}^\prime-m_{f_i})f_i \\
& + \overline{\chi}(-g_D \slashed{A}^\prime - m_\chi)\chi\,,
\end{split}\end{equation}
where $m_{f_i}$, $m_\chi$ and $M_{A^\prime}$ are the SM fermion, DM and dark photon masses, respectively, $F^{\mu \nu}$ and $F^{\prime\mu \nu}$ are the fields strength of the photon $A$ and of the dark photon $A^\prime$, $g_D$ is the coupling between the dark photon and  the dark sector, and $\varepsilon e$ the dark photon coupling with the standard fermion of charge $q_{f_i}$.\\

\begin{figure}[t!]
\begin{center}
\includegraphics[scale=0.3]{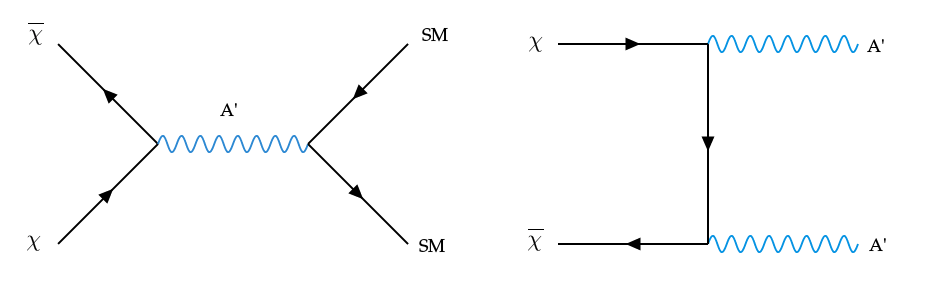}
\caption{Feynman diagrams for dark photon models.}
\label{diag}
\end{center}
\end{figure}

This model possesses two important channels for dark matter production, depending on the relation between the dark photon mass and the DM mass \cite{Alexander:2016aln}: for $m_\chi > M_{A^\prime}$, we have the $s$-channel annihilation into SM particles (left diagram in Fig.\ \ref{diag}) and on-shell production of two dark photons $A^\prime$ (right diagram in Fig.\ \ref{diag}), the latter being typically dominant. This scenario is called Secluded Dark Matter Model \cite{Pospelov:2007mp}. In the opposite case, $m_\chi < M_{A^\prime}$, we have $s$-channel annihilation producing a pair of (four) SM particles via the (off-shell) $A^\prime$ mediator (see Fig.\ \ref{diag}). In this work we will focus on the second case because  in this case the kinetic mixing parameter is a key parameter in the dark matter phenomenology, allowing to directly and straightforwardly explore the rich interplay of a multitude of independent searches for this specific class of dark photon models.\\

Having in mind the first diagram in Fig.\ref{diag} is straightforward to find the integrated amplitude squared which reads,
\begin{equation}\begin{split}\label{Amp2}
\int d\Omega |\mathcal{M}|^2 = & \frac{64\pi}{3} (\varepsilon e q_f g_D)^2 \left(1+ \frac{2m_f^2}{s}\right)\left(1+ \frac{2m_\chi^2}{s}\right) \frac{s^2}{(s-M_{A'}^2)^2+M_{A'}^2 \Gamma_{A'}^2},
\end{split}
\end{equation}where the total width of the dark photon is given by
\begin{equation}\begin{split}
\Gamma_{A'} =  \frac{M_{A'}}{4\pi} & \left[ \sum_i (\varepsilon e q_{f_i})^2 \left(1+\frac{2m_{f_i}^2}{M_{A'}^2}\right)\sqrt{1-\frac{4m_{f_i}^2}{M_{A'}^2}} + g_D^2 \left(1+\frac{2m_\chi^2}{M_{A'}^2}\right)\sqrt{1-\frac{4m_\chi^2}{M_{A'}^2}}  \right].
\end{split}
\end{equation}

These two expressions above will be important to understand the physical processes we will present further. 

\section{\label{sec:dm}Relic density}

The evolution of the number density of a species $i$ is set by the number of interactions per unit of volume and time, the rate $R(T)$. The corresponding Boltzmann equation reads
\begin{equation}
\dot{n}_i + 3H n_i = R(T)\,,
\end{equation}
where $H=\dot{a}/a$ is the Hubble rate, with $a$ the scale factor.

It is convenient to absorb the effect of the expansion of the Universe by defining the yield $Y_i = n_i/s$, where $s$ is the entropy density $s = S/a^3$. Since the entropy $S$ is dominated by the ultra-relativistic content, it holds that $s = 2\pi^2/45 g_s(T)T^3$, with $g_s$ the entropic relativistic degrees of freedom. The Boltzmann equation is therefore
\begin{equation}\label{evt}
\frac{dY_i}{dt} = \frac{R(T)}{s} - \frac{Y_i}{s}\frac{d\ln S}{dt}\,,
\end{equation}
where we see that entropy injection during dark matter production would decrease the dark matter yield. \\

Tracking the evolution in terms of temperature gives\footnote{Here we have used $dt = - \frac{1}{HT}dT + \frac{1}{3H}(d\ln S - d\ln g_s)$.}
\begin{equation}\label{evt}
\frac{dY_i}{dT} = - \tilde{g} \frac{R(T)}{sHT} - \frac{d\ln S}{dT}\left(Y_i - \frac{R(T)}{3Hs}\right),
\end{equation}
where $\tilde{g} \equiv \left(1+\frac{T}{3}\frac{d\ln g_s}{dT} \right)$. 
In our scenario, under entropy conservation we have 
\begin{equation}\label{chiev}
-\frac{T}{\tilde{g}} \frac{dY_\c}{dT} = 2 Y_\ap \frac{\G_{\ap \to \c\c}}{H}\left(1-\frac{Y_\c}{Y_\ap} \frac{n_\c \la \G_{\c\c \to \ap} \ra}{2\G_{\ap \to \c\c}}\right) + Y_f \frac{n_f \la \s v\ra}{H}\left(1-\frac{Y_\c^2}{Y_f^2} \right),
\end{equation}
where $\la \G_{\c\c \to \ap}\ra$ and $\la \s v\ra$ are the thermally averaged rates for inverse $A'$ decay and for $ff \leftrightarrow \chi \chi$, see below. 
Hereafter we will assume that the dark photons have already decoupled from the thermal bath, so that $Y_\ap \ll Y_f$ and we can drop the first term of the above equation. With Eq.\ (\ref{chiev}) at hand we have two possibilities for $Y_\chi$:

\begin{itemize}
\item {\it freeze-out}: $Y_f = Y_\c^{\rm eq}$. In the thermal freeze-out regime the dark matter was in thermal contact with SM particles, but eventually the expansion rate of the Universe equaled the interaction rate effectively preventing the dark matter particles to self-annihilate into SM particles, leading to freeze-out of the relic dark matter particle population. In this freeze-out scenario the yield  can lead to the correct dark matter relic density or not. We will discuss the case where dark matter abundance matches the one from freeze-out and the setup where the dark matter abundance is assisted by a late-time inflation episode. The latter will be addressed in Section \ref{sec:infl};

\item {\it freeze-in}: $Y_f \gg Y_\c$.  In the freeze-in case, the dark matter particles were never in equilibrium with fermions due to the weakness of their interactions with SM particles. It indeed {\it freezes-in} through the process $ff \to \c\c$.
\end{itemize}

We will now give a more quantitative description of these two processes. The rates for the freeze-out and the freeze-in regimes are respectively 
\begin{equation}\begin{split}
&R_{\rm FO}(T) = n_{\rm eq}^2 \la \s v \ra_{\rm ann} \left( 1 - \frac{Y_\c^2}{Y_{\rm eq}^2}\right) \\
& R_{\rm FI}(T) = n_f^2 \la \s v \ra_{\rm prod}
\end{split}\end{equation}

It turns out that we have
\begin{equation}\begin{split}\label{rate}
n_{\rm eq}^2 \la \s v\ra_{\rm ann} = n_f^2 \la \s v\ra_{\rm prod} & = \frac{T}{32(2\pi)^6} \int ds \sqrt{s} K_1\Big(\frac{\sqrt{s}}{T}\Big) \sqrt{1-\frac{4m_\c^2}{s}}\sqrt{1-\frac{4m_f^2}{s}} \int d\Omega |\M|^2,
\end{split}
\end{equation}where the number density of a species $i$ in this regime is
\begin{equation}
n_i = \frac{g_i}{2\pi^2} m_i^2 T K_2\left(\frac{m_i}{T}\right),
\end{equation}and $s$ is the Mandelstam variable and where $n_{\rm eq}$ stands for the equilibrium number density of dark matter. Furthermore, we highlight that we used the Maxwell-Boltzmann approximation, since we are interested in studying these processes in the non-relativistic regime of dark matter and fermions.\\

It is useful to have an analytic approximation for the thermally averaged annihilation cross section to also facilitate the understanding and interpretation of the bounds we will discuss further. In the limit $ m_f^2 \ll m_\chi^2 \ll M_{A'}^2$, i.e.\ when the dark matter annihilation to fermion pairs via the $s$-channel exchange of the dark photon is non-resonant,  $\la \s v \ra_{\rm ann} $ scales as
\begin{equation}
 \langle \sigma v \rangle_{\rm ann} \sim \frac{\left( g_D \varepsilon e\right)^2 m_\chi^2}{M_{A^\prime}^4}\,.
 \label{sigv}
\end{equation}
Therefore, if an experiment places a model independent bound on $\langle \sigma v \rangle_{\rm ann}$, we can interpret such limit in the $\varepsilon$ vs $M_{A^\prime}$ plane for a fixed dark matter mass. Furthermore, this constraint should weaken with the dark photon mass. This feature will clearly emerge in the following 
Figs.\ \ref{mdm001}-\ref{mdm01prospects}. Moreover, since the dark matter annihilation cross section is proportional to $1/s \int d\Omega |\mathcal{M}|^2$  defined in 
Eq.\ (\ref{Amp2}) and we can observe that the dark matter annihilation features a resonance when $M_{A^\prime}^2 \sim s \sim 4 m_{\chi}^2$, i.e.\ when $M_{A^\prime} \sim 2 m_{\chi}$, in the non-relativistic limit. This resonance regime is very important in model since most of the parameter space consistent with the existent limits, in the freeze-out scenario, lives near the $A^\prime$ resonance.\\

Now that we understood some limiting cases of the annihilation rate, we can go back to the discussion of the dark matter thermal relic abundance. In the freeze-out regime, we can rewrite Eq.\ (\ref{chiev}) as
\begin{equation}\begin{split}
\frac{dY_\c}{dT} &= - \frac{\tilde{g}}{sHT} n_{\rm eq}^2 \la \s v \ra_{\rm ann} \left(1-\frac{Y_\c^2}{Y_{\rm eq}^2}\right)
= M_{\rm Pl} \left(\frac{45}{\pi}\right)^{-1/2} g_*^{1/2} \la \s v \ra_{\rm ann} (Y_\c^2 - {Y}_{\rm eq}^2)\,
\end{split}\end{equation}
where we adopt the usual notation $g_*^{1/2} \equiv \tilde{g} g_s/\sqrt{g_e}$, with $g_e$ the sum of the degrees of freedom.\\

After freeze-out , we can take $Y_\c \gg Y_{\rm eq}$, giving a solution for the final yield $Y_0$
\begin{equation}\label{Yfo}
\frac{1}{Y_0} = \frac{1}{Y_f} + \frac{M_{\rm Pl}}{\sqrt{45/\pi}} \int_{T_0}^{T_{\rm fr}} dT g_*^{1/2} \la \s v \ra_{\rm ann}.
\end{equation}

Notice that in the freeze-out scenario we integrate from the freeze-out temperature $T_{\rm fr}$ to the temperature today $T_0$. The freeze-out temperature is found to be $T_{\rm fr} \sim m_{\chi}/10$ and it is derived by finding which temperature leads to a dark matter yield after freeze-out much larger than the yield in equilibrium. \\

$Y_f$ is defined as the yield of dark matter just after decoupling, $Y_f = Y_{\rm eq}(1+\delta)$, where for a good approximation $\delta = 1.5$ \cite{Gondolo:1990dk}. Since we are going to consider dark matter masses in the range $10-100$~MeV, the final states are just electrons.\\ 

For the freeze-in regime, Eq.\ (\ref{chiev}) reduces to
\begin{equation}\begin{split}\label{Yfi}
\frac{dY_\c}{dT} &\approx - \frac{\tilde{g}}{sHT} n_f^2 \la \s v \ra_{\rm prod} 
= - \frac{M_{Pl}}{(2\pi)^2} \left(\frac{45}{\pi}\right)^{3/2} \frac{\tilde{g}}{g_s \sqrt{g_e}} \frac{n_f^2 \la \s v \ra_{\rm prod}}{T^6},
\end{split}\end{equation}
which is easily integrated to give the final dark matter yield and relic density. In this case, we integrate from some high scale temperature, usually taken as the maximal temperature of radiation era, the reheating temperature, up to $T_0$. The sensitivity of the dark matter relic density on the reheating temperature in this case introduces an uncertainty to the dark matter physics. However, since most of the well motivated models provide a reheating temperature that is $T_{\rm rh}\gg $MeV, our results are not significantly affected by such an uncertainty. Interestingly, as pointed out in \cite{Maity:2018dgy}, it is possible to constraint dark matter physics through its dependence on the reheating temperature by using CMB observables.
\\

\begin{figure}[!t]
\begin{minipage}{0.5\linewidth}
\centering
\includegraphics[width=\textwidth]{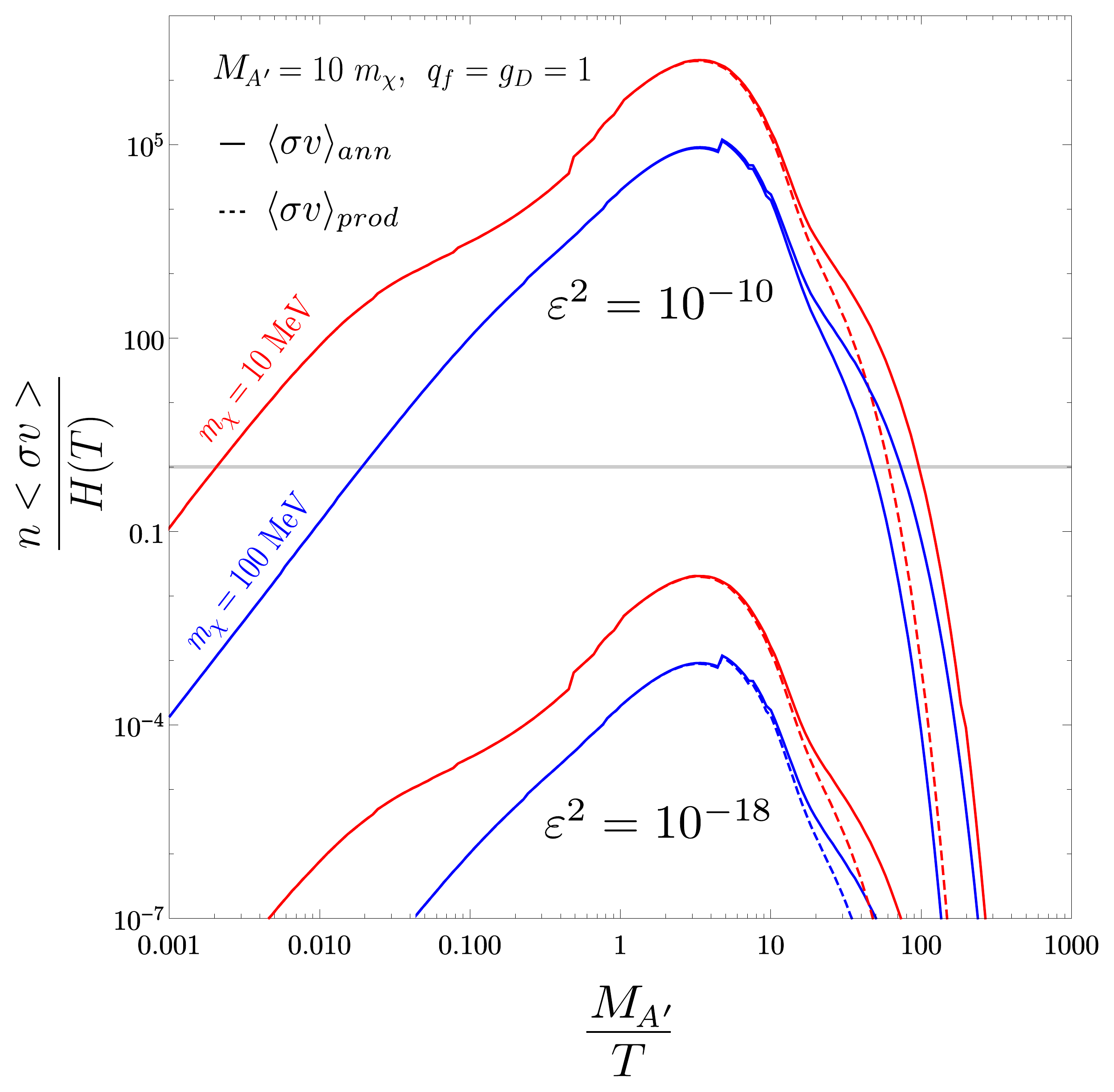}
\end{minipage}
\begin{minipage}{0.5\linewidth}
\centering
\includegraphics[width=\textwidth]{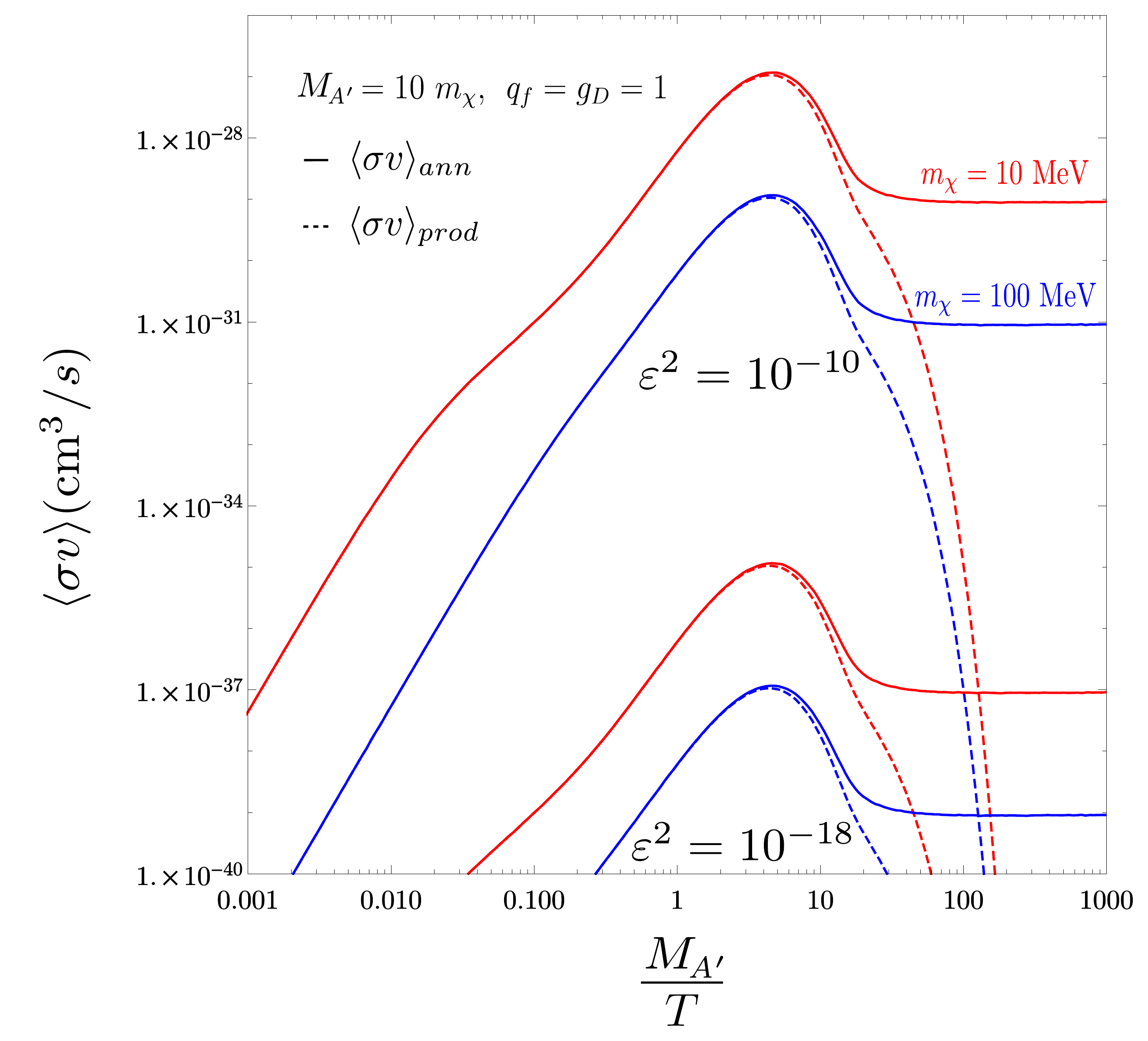}
\end{minipage}
\caption{{\it Left:} Ratio of the frequency of interactions to the frequency of expansion for a set of parameters of interest. Notice that for $\varepsilon^2 \lesssim 10^{-15}$ dark matter never reaches equilibrium with the fermions and could be produced via freeze-in. {\it Right:} Evolution of the thermally averaged annihilation and production cross sections.}
\label{sigmavs}
\end{figure}

The equilibrium condition between dark matter and fermions will dictate which mechanism will generate the dark matter relic density. Roughly speaking, if $n \la \s v \ra (T) < H(T)$, the expansion of the Universe is faster than the interactions and the sectors are considered thermally decoupled, otherwise they are in thermal contact. Therefore, if $n_{eq} \la \s v \ra_{\rm ann}/H > 1$, we have the freeze-out and if $n_f \la \s v \ra_{\rm prod}/H < 1$, we have the freeze-in. We illustrate, in the left panel of figure \ref{sigmavs}, the \textit{ratios $n \la \s v \ra (T) / H(T)$} as a function of the inverse temperature (rather, of the ratio $M_{A^\prime}/T$) for different choices of the annihilation rate and of the coupling $\varepsilon^2$.\\ 

Additionally, Fig.\ \ref{sigmavs} shows that the maximal pair-annihilation rate occurs in the pole region, when $s \sim M_\ap^2$. By using the Narrow Width Approximation (NWA), 
\begin{equation}
\int d\Omega |\M|^2 = \frac{A(s)}{(s-M^2)^2 + M^2 \Gamma^2} \rightarrow \frac{\pi}{M \Gamma} A(s) \delta (s-M^2) \,,
\end{equation}
we can estimate the point at which the maximum of the reaction rate is smaller than the expansion frequency. In the NWA approximation the rate is
\begin{equation}
R^{\rm NWA}(T) = \frac{M_\ap^4}{24\pi^3} \frac{K_1(x)}{x} \frac{1}{M_\ap (1/\G_{\ap \to \c\c} + 1/\G_{\ap \to ff} ) }.
\end{equation}
Since $\varepsilon \ll g_D$, $1/\G_{\ap \to ff} \gg 1/\G_{\ap \to \c\c}$ and therefore we will satisfy the out-of-equilibrium condition $R^{\rm NWA}(T)<H(T)$ roughly when 
\begin{equation}
\varepsilon^2 < 7.4 \times 10^{-16} \left(\frac{g_e}{10}\right)^{1/2} \left(\frac{M_\ap}{GeV}\right) \frac{r_f K_2(\sqrt{r_f}x) x^2 }{(1+ 2r_f)\sqrt{1-4r_f}K_1(x) },
\end{equation}
where $x \equiv M_\ap/T$ and $r_f \equiv m_f^2/M_\ap^2$. For $M_\ap \sim 100$~MeV, $g_e \sim 20$ and by considering electrons, $r_f \sim 10^{-5}$. Considering the maximum of the rate at $x \sim 3$, we can have freeze-in production for $\varepsilon^2 \lesssim 5 \times 10^{-15}$. \\

We highlight that in the freeze-out scenario, it is usual to work with the thermal annihilation cross section, which is $\la \s v \ra_{\rm ann} = R_{\rm FO}(T)/n_{\rm eq}^2$. Since both $R(T)$ and $n_{\rm eq}(T)$ decreases for $T \lesssim T_{\rm fr}$, the annihilation cross section becomes constant after the thermal decoupling. For the freeze-in scenario, we work instead with the production cross-section $\la \s v \ra_{\rm prod} = R_{\rm FI}(T)/n_f^2$. Since $n_f$ is still nearly constant after decoupling, $\la \s v \ra_{\rm prod}$ decreases after decoupling. We illustrate this in the {\it right panel} of figure \ref{sigmavs}. \\

One can thus integrate Eqs.~(\ref{Yfo}) and (\ref{Yfi}) in order to find the relic density in the freeze-out and freeze-in scenarios, respectively. In what follows we will discuss current and upcoming experimental bounds on the model, in particular from direct detection and indirect detection, and compare the constraints with the parameter space leading to the 
correct relic density.\\


\section{\label{sec:cmb}CMB bounds}

The results of the Planck satellite improved significantly the precision and resolution of the measurement of the anisotropies of the CMB spectrum, rendering the CMB bounds on dark matter annihilation very  competitive when compared to the standard relic density and indirect detection constraints. \\

If dark matter particles annihilate at early times, between the period of recombination and reionization, they could inject electromagnetic particles in the intergalactic medium. This could affect significantly the CMB power spectrum, for example, enlarging the surface of last scattering or increasing the electron ionization fraction, since this injected energy can ionize and heat the intergalactic medium.\\

The energy per time per volume deposited in the medium by a DM annihilation is given by
\begin{equation}
\frac{dE}{dt \,dV}=\rho_c^2 \Omega_\chi^2 (1+z)^6 P_{\rm ann}(z)\,,
\end{equation}
where $\rho_c$ is the critical density, $\Omega_\chi$ is the DM abundance, $z$ is the redshift, and the redshift-dependent parameter $P_{ann}(z)$ is the annihilation parameter, defined as
\begin{equation}
P_{\rm ann}(z) \equiv f(z)\frac{\left\langle  \sigma v  \right\rangle}{m_{\chi}}\,,
\end{equation}
which depends on the efficiency function $f(z)$ defined below, the thermal averaged annihilation cross section $\left\langle  \sigma v  \right\rangle$ and the dark matter mass $m_{\chi}$. The efficiency function $f(z)$ describes the relation between the deposited energy and the injected energy in the medium for a given redshift: 
\begin{equation}
\left.\frac{dE}{dt\,dV}\right|_{\rm dep}(z)= f(z) \left.\frac{dE}{dt\,dV}\right|_{\rm inj}(z)\,.
\end{equation}
In other words, $f(z)$ is the efficiency at which the deposited energy is actually injected into the medium as a function of the redshift. 
This function was carefully computed in \cite{Slatyer:2012yq}. Recently, it was demonstrated that 
with good precision the function $f(z)$ can be approximated to be independent of redshift,
being an effective efficiency factor $f_{\rm eff}$ 
\cite{Slatyer:2015kla, Slatyer:2015jla}. Thus, hereafter, we will refer to it as $f_{\rm eff}$.\\

This efficiency factor depends on the final state from dark matter annihilation as well. In our model the possible final states are electron-positron pairs and photons resulting from final state radiation.\\

In order to compute $f_{\rm eff}$ we need to obtain the numbers of electron-positron pairs and photons produced per dark matter annihilation as a function of energy, a quantity known as energy spectrum. We denote $dN/dE^\gamma$ and $dN/dE^{e^+}$ as the photon and positron energy spectrum respectively. The electron-positron yield is trivial and the photon production is in the approximation where the dark matter mass is much larger than the electron mass \cite{Fortin:2009rq}. A numerical calculations of these energy spectra with Pythia or PPPC4DM would furnish similar results \cite{Cirelli:2010xx,Sjostrand:2014zea}. With the energy spectra at hand we also need to account for the individual efficiency functions of the electron-positron pairs and photons. Such individual efficiency functions simply quantify how much these particles perturb the ionization history of the Universe as a function of energy. We label them as $f_{\rm eff}^\gamma$ and $f_{\rm eff}^{e^+}$.
Having the energy spectra at hand and these individual efficiency functions obtained using the code from Ref.\ \cite{Slatyer:2015jla}, we can compute the overall efficiency factor, $f_{\rm eff}$, by integrating over energy \cite{Slatyer:2015jla} as follows,

\begin{equation}
f_{\rm eff} = \frac{1}{2 m_{ \chi}}\int_0^{m_{\chi}}EdE\left(f_{\rm eff}^{\gamma}(E)\frac{dN}{dE^{\gamma}}\right. \nonumber + 2 \left. f_{\rm eff}^{e^+}(E)\frac{dN}{dE^{e^+}}\right),
\end{equation}
where the factor $2$ appear to account for electrons and positrons.\\ 

For now, using this approximation, 
$P_{\rm ann}$ can be rewritten as 
\begin{equation}
P_{\rm ann} \equiv f_{\rm eff} \frac{\langle \sigma v \rangle}{m_{\chi}}\,,
\label{eqPann}
\end{equation}
which is a $z$-independent quantity that currently is constrained by Planck to  
\cite{Ade:2015xua} 
\begin{equation}
P_{\rm ann} < 4.1 \times 10^{-28} \, {\rm cm}^3 \, {\rm s}^{-1} \, {\rm GeV}^{-1} \,,
\label{pann}
\end{equation}
which will be used to constrain the MeV dark matter in our dark photon model. With Eq.\ (\ref{eqPann}) at hand and the bound from Planck in Eq.\ (\ref{pann}) we can place a limit on the dark matter annihilation cross section as a function of the dark matter mass for a known $f_{\rm eff}$. We remind the reader that $f_{\rm eff}$ was obtained numerically using the routine provided in \cite{Slatyer:2015jla}. \\

At the end, we obtained the CMB bound displayed in Fig.\ \ref{sigmaVbound} as a red curve. To be more specific, we give two examples: our bound is $\left\langle  \sigma v  \right\rangle  < 5.18 \times 10^{-30} \, {\rm cm^3 \, s^{-1}}$ for $m_{\chi}=10$~MeV and $\left\langle \sigma v  \right\rangle < 5.65 \times 10^{-29} \,{\rm cm^3 \,s^{-1}}$ for $m_{\rm \chi}=100$~MeV.

\section{\label{sec:gam}Gamma-rays}

The last  decade has brought unprecedented progress in the area of GeV gamma-ray astronomy, improving the sensitivity to a dark matter signal by several orders of magnitude  \cite{Strong:1998ck,Knodlseder:2005yq,Aharonian:2006wh,Abeysekara:2017jxs,Archambault:2017wyh}.  However, in the MeV-GeV regime there has not been much progress and the e-ASTROGAM proposal is anticipated to fill this gap \cite{DeAngelis:2016slk}. \\

Interesting studies in the energy range relevant for MeV-GeV gamma-ray indirect detection have been presented e.g.\ in Refs.~\cite{Boehm:2003bt,Hooper:2003sh,Beacom:2004pe,Ahn:2005ck,Lawson:2007kp,Hooper:2007tu,Huh:2007zw,Kahn:2007ru,Boddy:2015efa,Queiroz:2014yna,Mambrini:2015sia,Bringmann:2016axu,Boddy:2016fds,Garcia-Cely:2016pse,Garcia-Cely:2016hsk,Brdar:2017wgy}. In this work, will focus on the e-ASTROGAM mission.   The e-ASTROGAM space observatory is projected to be comprised of a silicon tracker, a calorimeter and an anti-coincidence system, sensitive to photons in the energy range from $0.3$~MeV to $3$~GeV. 
That said, Ref.~\cite{Bartels:2017dpb} performed a dedicated sensitivity study of the e-ASTROGAM mission to MeV dark matter. There it was assumed that the systematic uncertainties in these future missions will be similar
to those pertinent to Fermi-LAT; the local dark matter density to be $0.4$ GeV/cm$^3$; the dark matter density profile to be modeled by a Navarro-Frenk-White halo \cite{Navarro:1995iw,Navarro:1996gj}; and the region of interest to be the Galactic center. For the case of dark matter annihilation into $e^+e^-$, they have included prompt photons resulting from final state radiation as well as secondary photons \cite{Aharonian:1981spy,Aharonian:2000iz}. Here we will consider the benchmark scenario described in \cite{Bartels:2017dpb} where final state radiation 
is the main component along with bremsstrahlung emission. \\

\begin{figure}[h!]
\centering
\includegraphics[scale=0.5]{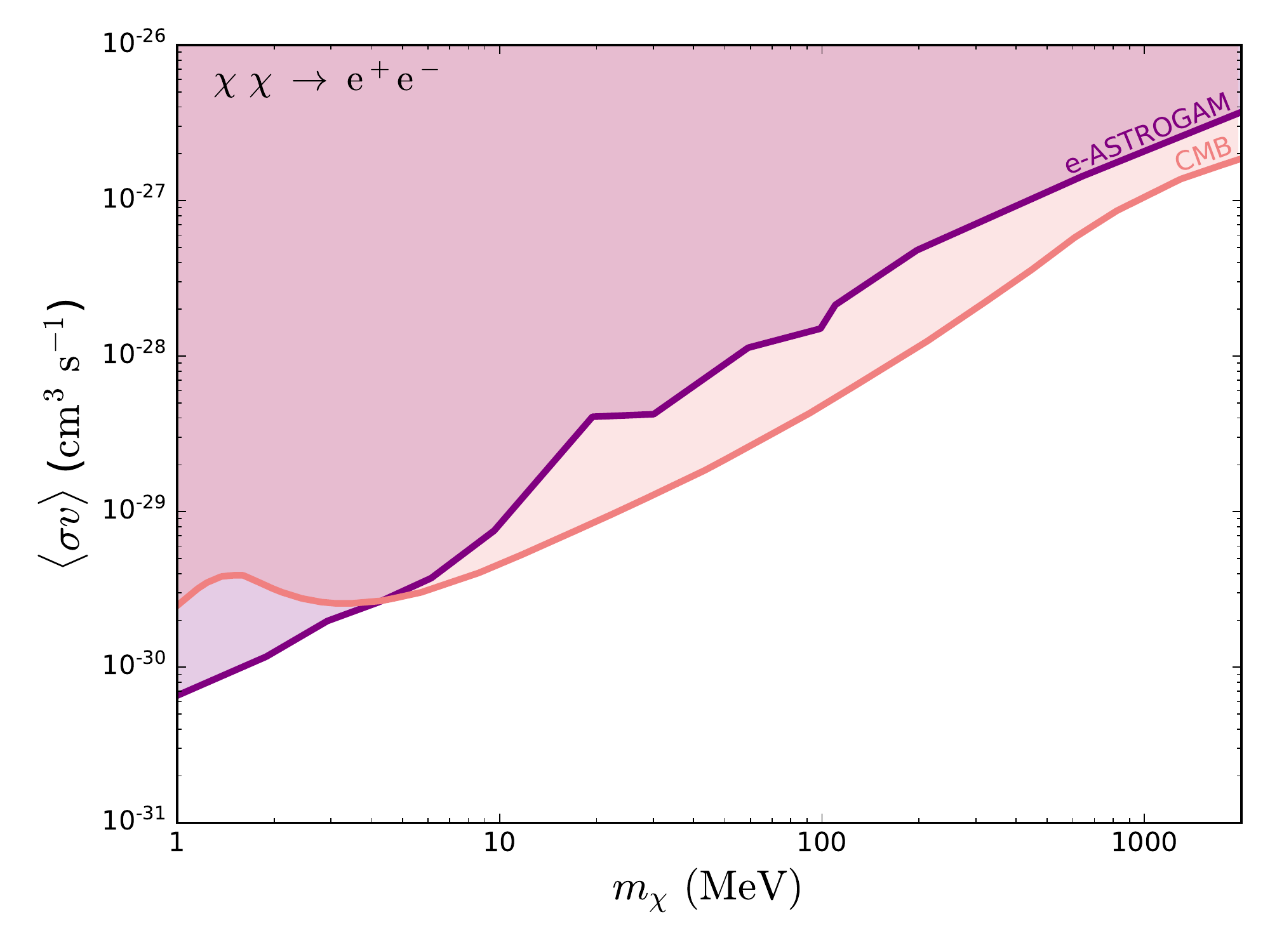}
\caption{Limits on cross section \textit{versus} dark matter mass from CMB bounds \cite{Slatyer:2015jla} and prospects from e-ASTROGAM \cite{Bartels:2017dpb} on the channel $\chi \chi$ $\rightarrow$ $e^+ e^-$. 
}
\label{sigmaVbound}
\end{figure}

In summary, the bound found in \cite{Bartels:2017dpb} is reproduced in Fig.\ \ref{sigmaVbound} with a purple line and compared with the CMB one derived in the previous section. It is clear that e-ASTROGAM can potentially discover dark matter for masses below $10$~MeV, while offering a complementary and important probe for larger dark matter masses. Thus, in Fig.\ \ref{mdm001}, where we exhibit the results for $m_{\rm \chi}=10$~MeV, e-ASTROGAM and CMB will constitute orthogonal and nearly equally competitive bounds.

\section{\label{sec:DD}Direct Detection}

Another promising way to discover MeV dark matter is via the observation of dark matter scatterings at nuclear targets or electrons \cite{Abe:2015eos,Angloher:2015ewa,Agnese:2015nto,Aprile:2016swn,Akerib:2016lao,Angloher:2016rji,Fu:2016ega,Akerib:2016vxi,Amole:2017dex,Aprile:2017ngb,Witte:2017qsy,Cui:2017nnn,Aprile:2017aty,Aprile:2017iyp,Aprile:2017aas,Aprile:2017kek,Aprile:2017yea}. In particular, the strongest limits on MeV dark matter stems from the XENON10 and -100 experiments, two-phase detectors that used ionization and scintillation to distinguish background from signal events \cite{Essig:2015cda,Essig:2017kqs}.\\

\begin{figure*}[h!]
\includegraphics[scale=0.37]{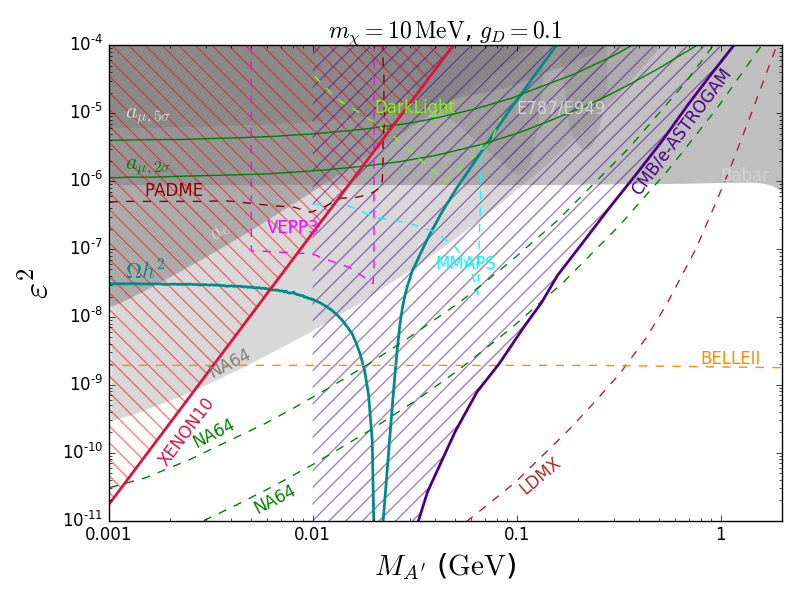}
\includegraphics[scale=0.37]{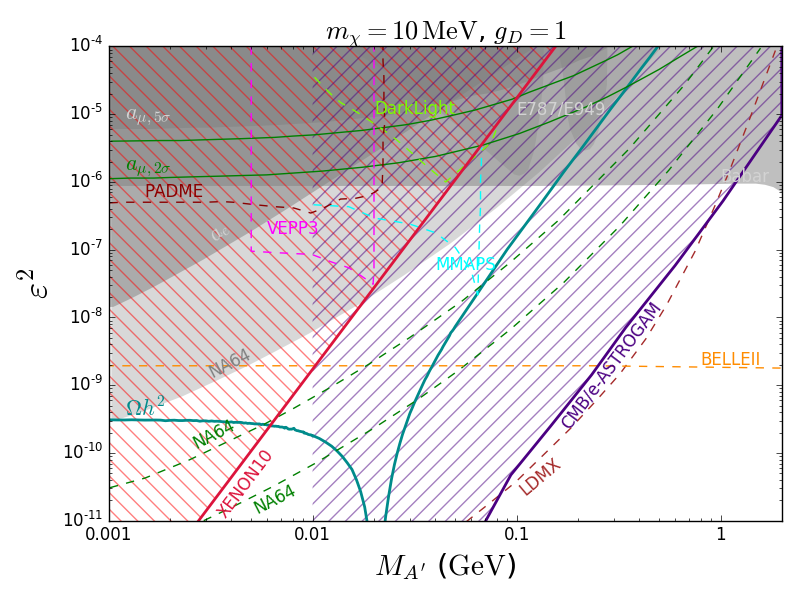}
\caption{
Bounds on the plane $\varepsilon^2$ \textit{versus} dark photon mass from CMB (see \cite{Slatyer:2015jla}) 
and direct detection (see \cite{Essig:2017kqs}) 
constraints on the channel $\chi \chi \rightarrow e^+ e^-$ (purple and red hatched regions, respectively) and relic abundance (turquoise lines) for DM mass $m_{\rm \chi}=10$~MeV and two different values for the dark coupling $g_{D}$, $g_{D}=0.1$ (left panel) and $g_{D}=1$ (right panel). We are comparing our results with the most recent dark photon searches (gray regions) and future prospects (colored regions).
}
\label{mdm001}
\end{figure*}

In the case of MeV dark matter, when a dark matter particle scatters off an electron it may ionize a xenon atom in the liquid phase. The recoiling electron can ionize other surrounding atoms if it has sufficient energy as well. An electric field then drifts the electrons to the xenon gas phase where a scintillation signal is produced, commonly referred to as the $S_2$ signal. This signal is proportional to the number of stripped electrons. Taking into account the specifics of the XENON10 and -100 detectors, bounds were placed on the dark matter-electron scattering cross section \cite{Essig:2015cda}. In particular, they found $\sigma_e < 4.5 \times 10^{-37}$ cm$^2$ for $m_{ \chi}=10$~MeV and $\sigma_e < 9 \times 10^{-39}$ cm$^2$ for $m_{\chi}=100$~MeV. Such bounds are displayed in red curves in Figs.\ \ref{mdm001}-\ref{mdm01current}.\\

Moreover, projected limits from the SuperCDMS collaboration using silicon with a 10 kg-year exposure were forecast to be $\sigma_e < 2.7 \times 10^{-43}$ cm$^2$ for $m_{\chi}=10$~MeV and $\sigma_e < 1.3 \times 10^{-42}$ cm$^2$ for $m_{\chi}=100$~MeV \cite{Essig:2015cda}. These limits will significantly reduce the viable parameter space of the model and we represent them with red curves in Fig.\ \ref{mdm01prospects}.\\

These model-independent bounds can be interpreted in terms of the dark photon model we consider knowing  that the dark matter-electron scattering cross section reads \cite{Essig:2015cda}
\begin{equation}
\sigma_e = \frac{16 \pi  \mu_{\chi e} \alpha \varepsilon^2 \alpha_D }{M_{A^\prime}^2 + \alpha^2 m_e^2}\,,
\end{equation}where $\alpha$ is the fine-structure electromagnetic constant, $\alpha_D = g_D/4\pi$, and $\mu_{\chi e}$ is the dark matter-electron reduced mass. For a fixed dark matter mass $\mu_{\chi e}$ is determined and one can thus translate the experimental bound on $\sigma_e$ into a bound on the $\varepsilon$ vs $M_{A^\prime}$ plane. \\

We now move to the discussion of dark matter production via freeze-out and freeze-in on the same parameter space used before.



\section{\label{sec:alt}Dark Matter Production}

The evidence for dark matter in our Universe is irrefutable, but the production mechanism for the dark matter in the early Universe remains utterly mysterious. In a nutshell, dark matter particles can be produced either in thermal equilibrium or not. The former is known at thermal freeze-out. In the thermal freeze-out regime the dark matter was in thermal contact with SM particles, but eventually the expansion rate of the Universe equaled the interaction rate effectively preventing the dark matter particles to self-annihilate into SM particles, leading to freeze-out of the relic dark matter particle population. If dark matter particles never attain equilibrium in the early Universe, there is no unique way to non-thermally generate the observed dark matter abundance \cite{Scherrer:1985zt,Gondolo:1990dk,Kamionkowski:1990ni}.  
One possibility is that processes dump out-of-equilibrium dark matter particles in the early Universe, letting a slowly-growing population of particles to accrete, eventually, to the observed abundance -- a process dubbed freeze-in \cite{Hall:2009bx,Blennow:2013jba,Klasen:2013ypa,Shakya:2015xnx,Bernal:2017kxu}.

\subsection{Freeze-out}

In Fig.\ \ref{mdm001} we summarize the results for $m_{\chi}=10$~MeV. In the left (right) panel we exhibit the limits for $g_D=0.1$ ($g_D=1$). The gray regions represent current limits from BaBar \cite{Lees:2017lec}, muon $g-2$ \cite{Davoudiasl:2014kua}, E787/E949 \citep{Adler:2004hp,Artamonov:2008qb,Essig:2013vha} and NA64 \cite{Banerjee:2017hhz} ranging from accelerators to colliders as reviewed in \cite{Alexander:2016aln,Battaglieri:2017aum}. The colored dashed lines account for projected sensitivities of a multitude of experiments such as NA64, LDMX, BELLE II etc \cite{Wojtsekhowski:2012zq,Raggi:2014zpa,Raggi:2015gza,Balewski:2014pxa,Battaglieri:2017aum,Wojtsekhowski:2017ijn}. The red curve is the current XENON exclusion limit, whereas the purple curve delimits the current CMB bound as well as the forecast e-ASTROGAM sensitivity. The region of parameter space that yields the correct dark matter relic density is demarcated by a green solid curve. One can easily notice that the relic density curve is completely immersed in the exclusion region of the e-ASTROGAM/CMB probes for $g_D=0.1$. Therefore, there is no room for a $10$~MeV Dirac fermion dark matter candidate for $g_D=0.1$ and a small space for $g_D=1$ that will be probed in the next generation of NA64 and LDMX whose relic density stems entirely from the thermal freeze-out process.\\

In  Fig.\ \ref{mdm01current} we display the results for $m_{\chi}=100$~MeV with $g_D=0.1$ (left panel) and $g_D=1$ (right panel). In this figure we introduce a dilution parameter $\Delta$, to be discussed in more detail further below. The important point is that $\Delta=1$ corresponds to the case where the dark matter relic density arises only from the thermal production of dark matter. This scenario is represented by the green solid curve.  That said, we can conclude from the left panel that a $100$~MeV dark matter with $g_D=0.1$ is excluded, whereas $g_D=1$ has a small viable region for $M_{A^\prime} < 10^{-1}$~GeV. In summary, only for $g_{D}=1$ one can accommodate an MeV dark matter particles in the dark photon portal without need of non-standard cosmology. We will now discuss the setup where $\Delta > 1$.

\subsection{\label{sec:infl}Freeze-out followed by late-time inflation }

It is possible to bring models with an over-abundant thermal relic into accord with observations by invoking  non-standard cosmology. 
If some beyond the SM field had driven a phase transition while dominating the energy density of the Universe for a short period of time prior to big bang nucleosynthesis (BBN), it would provoke an acceptable late-time inflation \cite{Davoudiasl:2015vba}. Since such a phenomenon could inject a significant amount of entropy into the thermal bath, any already decoupled species $i$ of yield $Y_i = n_i/s$ would undergo a dilution:
\begin{equation}
Y_i^A = \frac{Y_i^B}{\Delta}.
\end{equation}
Here $Y_i^B$ denotes the yield before dilution, which would be set by the freeze-out for instance, and $Y_i^A$ denotes the diluted yield, after the dilution process. 
Therefore, the parameter $\Delta$ quantifies the effect of late-time inflation 
into the abundance of the dark matter: an over-abundant dark matter density is simply diluted by this factor $\Delta$. \\

Actually, there are many ways to accomplish an entropy injection episode, and several realizations have been discussed in the literature \cite{Hooper:2011aj,DiBari:2013dna,Kelso:2013paa,Kelso:2013nwa,Baer:2014eja,Queiroz:2014ara,Allahverdi:2014bva,Merle:2015oja,Aoki:2015nza,Okada:2015kkj,Kane:2015jia,Kim:2016spf,Aparicio:2016qqb,DEramo:2017gpl,Bramante:2017obj,Dimastrogiovanni:2017tvd,Allahverdi:2017sks,Baur:2017stq,Hoof:2017ibo}. We will remain agnostic about the origin of this entropy injection episode and simply assume it existed after the dark matter freeze-out but before BBN and quantify its consequences in the context of complementary probes for dark matter. That said, we cannot discuss the dilution factor for $10$~MeV dark matter because the freeze-out occurs at BBN.  For $100$~MeV dark matter, however, we can discuss it since the freeze-out occurs before BBN, allowing the entropy injection episode to take place. \\



\begin{figure*}[t!]
\includegraphics[scale=0.37]{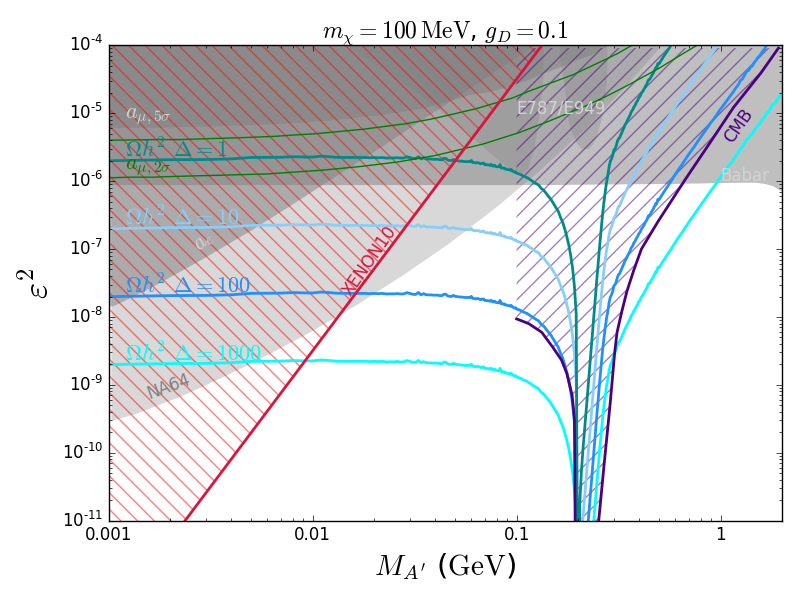}
\includegraphics[scale=0.37]{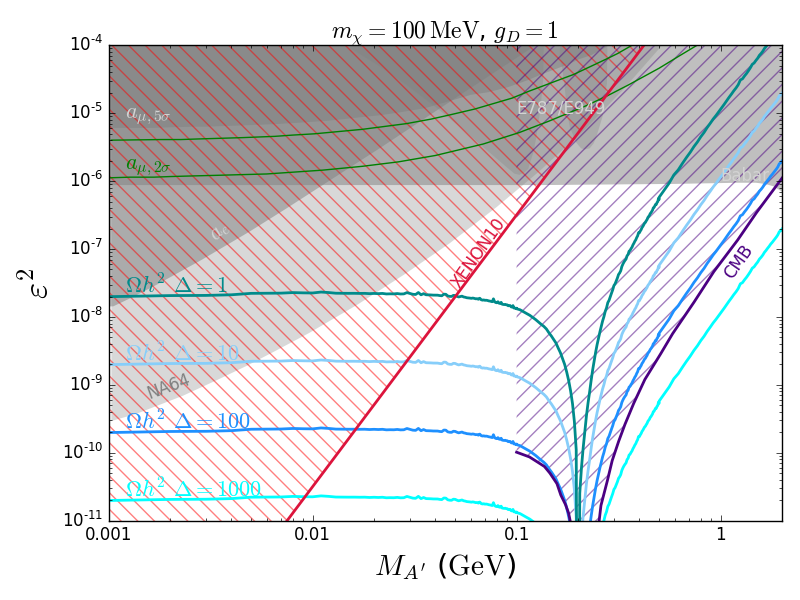}
\caption{Bounds on the plane $\varepsilon^2$ \textit{versus} dark photon mass from CMB (see \cite{Slatyer:2015jla}) and direct detection (see \cite{Essig:2017kqs}) constraints on the channel $\chi\chi$ $\rightarrow$ $e^+ e^-$ (purple and red hatched regions, respectively) and relic abundance (turquoise lines) for DM mass $m_{\rm \chi}=100$~MeV, for two different values for the dark coupling $g_{D}$, $g_{D}=0.1$ (left panel) and $g_{D}=1$ (right panel). We are comparing our results with the most recent  dark photon searches (gray regions). Here $\Delta$ is a dilution factor resulted from late-time inflation needed to suppress the dark matter relic density, which was initially overclosing the Universe.}
\label{mdm01current}
\end{figure*}

Anyways, we remind the reader that the annihilation cross section goes as $\langle  \sigma v  \rangle \sim \varepsilon^{2} m_{\chi}^{2}/ M_{A^\prime}^{4}$ and the abundance grows with $1/\langle\sigma v\rangle$. Hence, the smaller $\varepsilon$ the larger the dark matter abundance. Due to the existence of stringent limits from direct detection, indirect and collider experiments, we will be forced to live in a region of parameter space which $\varepsilon$ is very small, leading to an over-abundant dark matter candidate. For this reason we need this late-time inflation episode, i.e.\ the dilution factor $\Delta$, since it suppresses the dark matter relic density bringing it down to the correct value. Moreover, the smaller $\varepsilon$ the larger $\Delta$ needed to reproduce the correct relic density. This effect is clearly visible in Fig.\ \ref{mdm01current}. \\

In the Fig.\ \ref{mdm01current} we present the result for $m_{\chi}=100$~MeV with $g_{D}=0.1$ (left panel) and $g_{D}=1$ (right-panel). We preserve the same color scheme of the previous figure, where the gray area represent the existing limits on the model.  From the left panel of Fig.\ \ref{mdm01current} we see that $\Delta \gtrsim 10$ is needed to find a region of parameter space yielding the correct relic density while simultaneously obeying experimental limits. It is interesting to see that accelerators provide a complementary probe for MeV dark matter. In particular, for $\Delta \gtrsim 100$, accelerators are the most promising detection method. \\

For $g_{D}=1$ (right panel of Fig.\ \ref{mdm01current}) the complementarity among all these searches is fascinating with direct detection being very restrictive for $M_{A^\prime} < 50$~MeV, accelerators for $\Delta \gtrsim 100$, and indirect detection for $M_{A^\prime} > 100$~MeV. Notice that a $100$~MeV dark matter is perfectly consistent with all existing bounds with no need for non-standard cosmology. \\

In summary, in light of existing constraints only for a small region of parameter space can one accommodate an MeV dark matter candidate based on thermal production of dark matter and standard cosmology. The departure from a standard cosmology opens up a lot the viable parameter space of the model allowing both low and large dark matter masses to accommodate MeV dark matter.\\

\begin{figure*}[t!]
\includegraphics[scale=0.37]{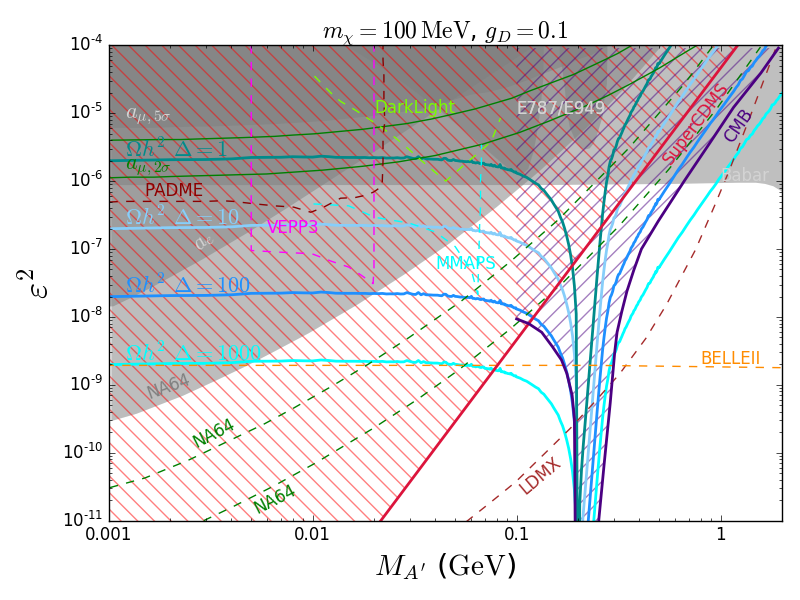}
\includegraphics[scale=0.37]{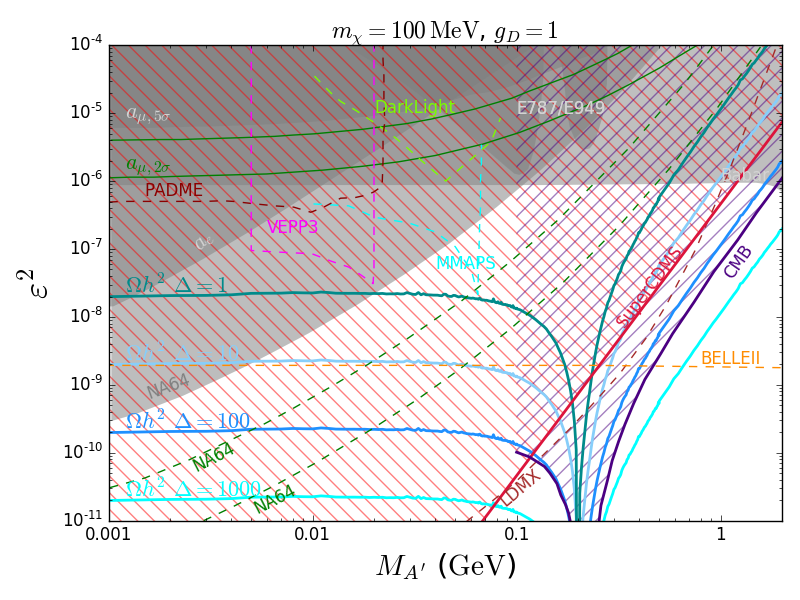}
\caption{Bounds on the plane $\varepsilon^2$ \textit{versus} dark photon mass from CMB constraints (see \cite{Slatyer:2015jla}) and direct detection prospects (see  \cite{Essig:2015cda}) on the channel $\chi \chi$ $\rightarrow$ $e^+ e^-$ (purple and red hatched regions, respectively) and relic abundance (turquoise lines) for DM mass $m_{\chi}=100$~MeV, for two different values for the dark coupling $g_{D}$, $g_{D}=0.1$ (left panel) and $g_{D}=1$ (right panel). We are comparing our results with the most recent searches on dark photons (gray regions) and future prospects (dashed colored lines). Again, $\Delta$ is a dilution factor resulted from late-time inflation needed to suppress the dark matter relic density, which was initially overclosing the Universe. 
}
\label{mdm01prospects}
\end{figure*}

 It is nonetheless important to have in mind prospects for MeV dark matter in the dark photon portal. To illustrate that, we display in Fig.\ \ref{mdm01prospects} projected limits from direct detection assuming the SuperCDMS setup \cite{Agnese:2016cpb,Agnese:2017njq} following the receipt given in \cite{Essig:2015cda}. We notice that a SuperCDMS-like detector is very important and might detect MeV dark matter, covering a large region of the parameter space of the model where a correct relic density is achieved either via thermal production or late-time inflation. \\

Moreover, we have introduced the projected limits from colliders and accelerators (NA64, LDMX among others) \cite{Alexander:2016aln}. It is exciting to see that such experiments can almost fully test the model, regardless of the dark matter production mechanism assumed (freeze-out and/or late-time inflation). 

\subsection{Freeze-in}

In this section we present the case of dark matter freeze-in production. We emphasize that in this mechanism the dark matter particle never reaches equilibrium with SM particles. In order to successfully achieve the DM production via freeze-in, the kinetic mixing parameter has to be finely tuned to small values. Therefore direct and indirect limits we discussed previously are no longer relevant. \\

\begin{figure*}[t!]
\includegraphics[scale=0.37]{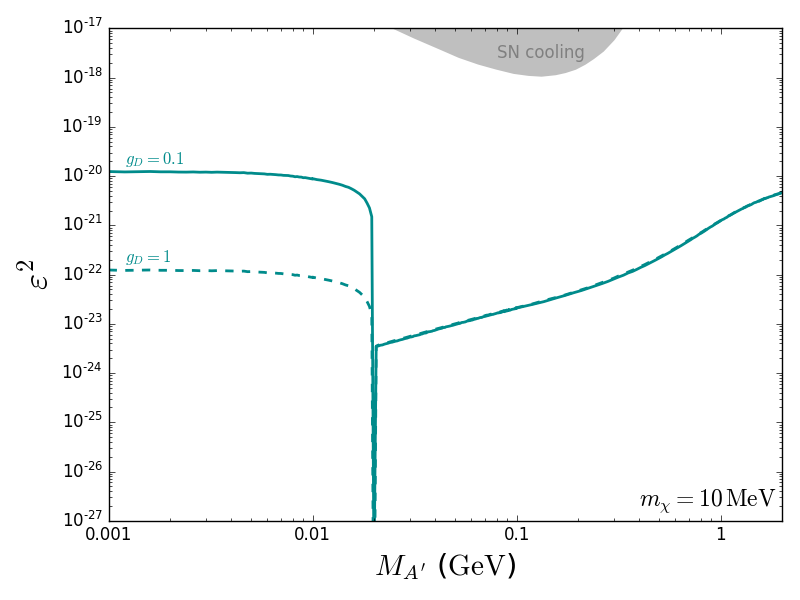}
\includegraphics[scale=0.37]{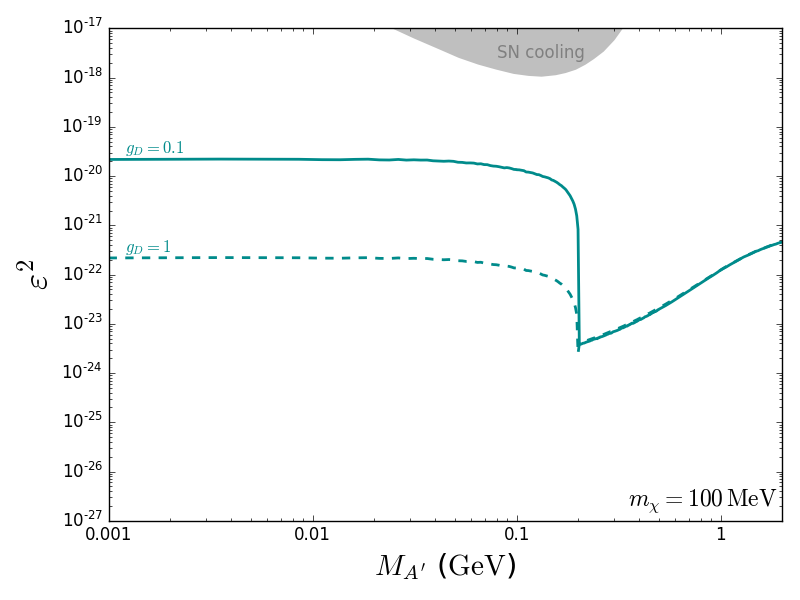}
\caption{Freeze-in abundance constraint on the plane $\varepsilon^2$ \textit{versus} dark photon mass (turquoise lines) for DM mass, $m_{\chi }=10$~MeV (left panel) and $m_{\chi}=100$~MeV (right panel), for two different values for the dark coupling $g_{D}$, $g_{D}=0.1$ (continuous lines) and $g_{D}=1$ (dashed line). We are comparing our results with the most recent dark photon searches (gray regions).}
\label{freezeinfig}
\end{figure*}

Any weakly interacting light species that can be produced in a supernova event  can potentially affect the energy loss and thus the luminosity of a supernova episode \cite{Turner:1987by}. Since the neutrino observation from SN1987A \cite{Hirata:1987hu,Bionta:1987qt} strong limits have been imposed on new light particles such as axions  and dark photons \cite{Dreiner:2013mua,Kazanas:2014mca,Mahoney:2017jqk,smolinsky_dark_2017}. In our case, these new dark photons could be emitted in the channels like $p+p \rightarrow p+p+A^\prime$ and $p+n \rightarrow p+n+A^\prime$ via bremsstrahlung and for the second case via pion emission too. This emission alters the energy loss of the supernova, which can be expressed in terms of the luminosity in the emitted light particle. The maximum energy loss $\epsilon_A$ permitted by the SN1987A observation is given by \cite{Raffelt:1996wa},
\begin{equation}
\epsilon_A=\frac{L_A}{M}\sim 10^{19} \, \frac{\mathrm{erg}}{\mathrm{g.s}}\,,
\end{equation}
where $M$ is the supernova mass and and $L_A$ its luminosity. This constraint imposes a lower limit on the $\varepsilon$ parameter. However, for large $\varepsilon$ the dark photon could decay before having left the supernova core or get trapped and thermalize, which effectively produces an upper limit on $\varepsilon$ for which constraints are effective \cite{Dreiner:2013mua}. In short, we find that supernova physics does not constrain the relic density curve of our model as one can observe in Fig.\ \ref{freezeinfig}. BBN constraints arising due to the cascade reaction induced by a very long lived dark photon are not directly applicable to our model either, because our dark photon decays into dark matter \cite{Fradette:2014sza}.\\

Hence, one can successfully produce MeV DM via  freeze-in in the dark photon portal escaping most phenomenological constraints, unlike the thermal equilibrium case discussed above. This fact is clearly visible in Fig.\ \ref{freezeinfig} where the curves that delimit the parameter space which yields the correct relic density are free from constraints, for either dark matter masses, $10$~MeV and $100$~MeV. 

\section{\label{sec:concl}Conclusions} 

In this study, we have addressed MeV dark matter complementary in the context of the dark photon portal. The particle dark matter candidate was assumed to be a Dirac fermion that interacts with  SM particles via a kinetic mixing term between the $U(1)$ gauge fields. We have investigated the DM production via freeze-out solely and concluded that DM masses of $10$~MeV exhibit a small region of parameter space in accordance with existing limits; in the future a SuperCDMS-like experiment and projected bounds stemming from NA64 and LDMX detectors are potentially capable of probing entirely the scenario.\\

We showed, however, that a late-time inflation episode opens up the parameter space accommodating $100$~MeV DM particles in agreement with stringent limits from Planck and XENON experiments. Orthogonal probes provided by accelerators can test regions of the parameter space otherwise inaccessible to direct and indirect detection experiments, thus highlighting the importance of complementary searches for MeV dark matter. Projected limits in the context of indirect detection (e-ASTROGAM), colliders and accelerators (NA64 and LDMX for example), and direct detection (SuperCDMS) have also been introduced to show that such experiments are capable to test almost the entire parameter space of the model.\\

Lastly, we have studied the case of DM production via freeze-in in the dark photon portal to show that due to the small couplings involved, this scenario is basically free from constraints and offers a viable framework to host an MeV dark matter candidate.

\section*{Acknowledgments}
The authors thank  Johannes Bl\"umlein, Pierre Fayet, Sergei Gninenko, Felix Kahlhoefer and Pankaj Saha for fruitful comments. FSQ is grateful to Yann Mambrini and Abdelhak Djouadi from Orsay-LPT for the hospitality during the early stages of this project. FSQ acknowledges support from MEC, UFRN and ICTP-SAIFR FAPESP grant 2016/01343-7. SP is partly supported by the U.S.\ Department of Energy grant number de-sc0010107. CS acknowledges support from the CAPES/PDSE Process 88881.134759/2016-01.
MD acknowledges support from the Brazilian PhD program ``Ci\^encias sem Fronteiras''-CNPQ Process No.\ 202055/2015-9. 
WR is supported by the DFG with grant RO 2516/6-1 in the
Heisenberg program.

\bibliographystyle{JHEPfixed}
\bibliography{darkmatter}

\providecommand{\href}[2]{#2}\begingroup\raggedright\begin{thebibliography}{100}

\bibitem{Ade:2015xua}
{\bf Planck} Collaboration, P.~A.~R. Ade {\em et.~al.}, {\it {Planck 2015
  results. XIII. Cosmological parameters}},
  \href{http://xxx.lanl.gov/abs/1502.01589}{{\tt 1502.01589}}.

\bibitem{FAYET1990743}
P.~Fayet, {\it Extra u(1)'s and new forces},  {\em Nuclear Physics B} {\bf 347}
  (1990), no.~3 743 -- 768.

\bibitem{Ackerman:mha}
L.~Ackerman, M.~R. Buckley, S.~M. Carroll, and M.~Kamionkowski, {\it {Dark
  Matter and Dark Radiation}},  {\em Phys. Rev.} {\bf D79} (2009) 023519,
  [\href{http://xxx.lanl.gov/abs/0810.5126}{{\tt 0810.5126}}]. [,277(2008)].

\bibitem{Knapen:2017xzo}
S.~Knapen, T.~Lin, and K.~M. Zurek, {\it {Light Dark Matter: Models and
  Constraints}},  \href{http://xxx.lanl.gov/abs/1709.07882}{{\tt 1709.07882}}.

\bibitem{Fayet:1980rr}
P.~Fayet, {\it {On the Search for a New Spin 1 Boson}},  {\em Nucl. Phys.} {\bf
  B187} (1981) 184--204.

\bibitem{Fayet:1980ss}
P.~Fayet, {\it {PARITY VIOLATION EFFECTS INDUCED BY A NEW GAUGE BOSON}},  {\em
  Phys. Lett.} {\bf 96B} (1980) 83--88.

\bibitem{Fayet:2006xd}
P.~Fayet, {\it {U-boson detectability, and Light Dark Matter}},
  \href{http://xxx.lanl.gov/abs/hep-ph/0607094}{{\tt hep-ph/0607094}}.

\bibitem{An:2014twa}
H.~An, M.~Pospelov, J.~Pradler, and A.~Ritz, {\it {Direct Detection Constraints
  on Dark Photon Dark Matter}},  {\em Phys. Lett.} {\bf B747} (2015) 331--338,
  [\href{http://xxx.lanl.gov/abs/1412.8378}{{\tt 1412.8378}}].

\bibitem{Lees:2014xha}
{\bf BaBar} Collaboration, J.~P. Lees {\em et.~al.}, {\it {Search for a Dark
  Photon in $e^+e^-$ Collisions at BaBar}},  {\em Phys. Rev. Lett.} {\bf 113}
  (2014), no.~20 201801, [\href{http://xxx.lanl.gov/abs/1406.2980}{{\tt
  1406.2980}}].

\bibitem{Batley:2015lha}
{\bf NA48/2} Collaboration, J.~R. Batley {\em et.~al.}, {\it {Search for the
  dark photon in $\pi^0$ decays}},  {\em Phys. Lett.} {\bf B746} (2015)
  178--185, [\href{http://xxx.lanl.gov/abs/1504.00607}{{\tt 1504.00607}}].

\bibitem{Aguilar-Arevalo:2016zop}
{\bf DAMIC} Collaboration, A.~Aguilar-Arevalo {\em et.~al.}, {\it {First
  Direct-Detection Constraints on eV-Scale Hidden-Photon Dark Matter with DAMIC
  at SNOLAB}},  {\em Phys. Rev. Lett.} {\bf 118} (2017), no.~14 141803,
  [\href{http://xxx.lanl.gov/abs/1611.03066}{{\tt 1611.03066}}].

\bibitem{Fayet:2016nyc}
P.~Fayet, {\it {The light $U$ boson as the mediator of a new force, coupled to
  a combination of $Q,\ B,\ L$ and dark matter}},  {\em Eur. Phys. J.} {\bf
  C77} (2017), no.~1 53, [\href{http://xxx.lanl.gov/abs/1611.05357}{{\tt
  1611.05357}}].

\bibitem{Aaij:2017rft}
{\bf LHCb} Collaboration, R.~Aaij {\em et.~al.}, {\it {Search for dark photons
  produced in 13 TeV $pp$ collisions}},
  \href{http://xxx.lanl.gov/abs/1710.02867}{{\tt 1710.02867}}.

\bibitem{Ablikim:2017aab}
{\bf BESIII} Collaboration, M.~Ablikim {\em et.~al.}, {\it {Dark Photon Search
  in the Mass Range Between 1.5 and 3.4 GeV/$c^2$}},  {\em Phys. Lett.} {\bf
  B774} (2017) 252--257, [\href{http://xxx.lanl.gov/abs/1705.04265}{{\tt
  1705.04265}}].

\bibitem{Lees:2017lec}
{\bf BaBar} Collaboration, J.~P. Lees {\em et.~al.}, {\it {Search for Invisible
  Decays of a Dark Photon Produced in ${e}^{+}{e}^{-}$ Collisions at BaBar}},
  {\em Phys. Rev. Lett.} {\bf 119} (2017), no.~13 131804,
  [\href{http://xxx.lanl.gov/abs/1702.03327}{{\tt 1702.03327}}].

\bibitem{Choi:2017kzp}
K.-Y. Choi, K.~Kadota, and I.~Park, {\it {Constraining dark photon model with
  dark matter from CMB spectral distortions}},  {\em Phys. Lett.} {\bf B771}
  (2017) 162--167, [\href{http://xxx.lanl.gov/abs/1701.01221}{{\tt
  1701.01221}}].

\bibitem{Arcadi:2017kky}
G.~Arcadi, M.~Dutra, P.~Ghosh, M.~Lindner, Y.~Mambrini, M.~Pierre, S.~Profumo,
  and F.~S. Queiroz, {\it {The Waning of the WIMP? A Review of Models,
  Searches, and Constraints}},  \href{http://xxx.lanl.gov/abs/1703.07364}{{\tt
  1703.07364}}.

\bibitem{Kile:2009nn}
J.~Kile and A.~Soni, {\it {Hidden MeV-Scale Dark Matter in Neutrino
  Detectors}},  {\em Phys. Rev.} {\bf D80} (2009) 115017,
  [\href{http://xxx.lanl.gov/abs/0908.3892}{{\tt 0908.3892}}].

\bibitem{Izaguirre:2013uxa}
E.~Izaguirre, G.~Krnjaic, P.~Schuster, and N.~Toro, {\it {New Electron
  Beam-Dump Experiments to Search for MeV to few-GeV Dark Matter}},  {\em Phys.
  Rev.} {\bf D88} (2013) 114015, [\href{http://xxx.lanl.gov/abs/1307.6554}{{\tt
  1307.6554}}].

\bibitem{Izaguirre:2015pva}
E.~Izaguirre, G.~Krnjaic, and M.~Pospelov, {\it {MeV-Scale Dark Matter Deep
  Underground}},  {\em Phys. Rev.} {\bf D92} (2015), no.~9 095014,
  [\href{http://xxx.lanl.gov/abs/1507.02681}{{\tt 1507.02681}}].

\bibitem{Hochberg:2017wce}
Y.~Hochberg, Y.~Kahn, M.~Lisanti, K.~M. Zurek, A.~Grushin, R.~Ilan, S.~M.
  Griffin, Z.-F. Liu, and S.~F. Weber, {\it {Detection of sub-MeV Dark Matter
  with Three-Dimensional Dirac Materials}},
  \href{http://xxx.lanl.gov/abs/1708.08929}{{\tt 1708.08929}}.

\bibitem{Mei:2017etc}
D.~M. Mei, G.~J. Wang, H.~Mei, G.~Yang, J.~Liu, M.~Wagner, R.~Panth, K.~Kooi,
  Y.~Y. Li, and W.~Z. Wei, {\it {Direct Detection of MeV-Scale Dark Matter
  Utilizing Germanium Internal Amplification for the Charge Created by the
  Ionization of Impurities}},  \href{http://xxx.lanl.gov/abs/1708.06594}{{\tt
  1708.06594}}.

\bibitem{An:2017ojc}
H.~An, M.~Pospelov, J.~Pradler, and A.~Ritz, {\it {Direct Detection of
  MeV-scale Dark Matter via Solar Reflection}},
  \href{http://xxx.lanl.gov/abs/1708.03642}{{\tt 1708.03642}}.

\bibitem{Darme:2017glc}
L.~Darm\'e, S.~Rao, and L.~Roszkowski, {\it {Light dark Higgs boson in minimal
  sub-GeV dark matter scenarios}},
  \href{http://xxx.lanl.gov/abs/1710.08430}{{\tt 1710.08430}}.

\bibitem{Boudaud:2016mos}
M.~Boudaud, J.~Lavalle, and P.~Salati, {\it {Novel cosmic-ray electron and
  positron constraints on MeV dark matter particles}},  {\em Phys. Rev. Lett.}
  {\bf 119} (2017), no.~2 021103,
  [\href{http://xxx.lanl.gov/abs/1612.07698}{{\tt 1612.07698}}].

\bibitem{Choudhury:2017osc}
D.~Choudhury and D.~Sachdeva, {\it {Model Independent analysis of MeV scale
  dark matter}},  \href{http://xxx.lanl.gov/abs/1711.03691}{{\tt 1711.03691}}.

\bibitem{Bertuzzo:2017lwt}
E.~Bertuzzo, C.~J. Caniu~Barros, and G.~Grilli~di Cortona, {\it {MeV Dark
  Matter: Model Independent Bounds}},  {\em JHEP} {\bf 09} (2017) 116,
  [\href{http://xxx.lanl.gov/abs/1707.00725}{{\tt 1707.00725}}].

\bibitem{Arhrib:2015dez}
A.~Arhrib, C.~Boehm, E.~Ma, and T.-C. Yuan, {\it {Radiative Model of Neutrino
  Mass with Neutrino Interacting MeV Dark Matter}},  {\em JCAP} {\bf 1604}
  (2016), no.~04 049, [\href{http://xxx.lanl.gov/abs/1512.08796}{{\tt
  1512.08796}}].

\bibitem{Huang:2013zga}
J.~Huang and A.~E. Nelson, {\it {MeV dark matter in the 3+1+1 model}},  {\em
  Phys. Rev.} {\bf D88} (2013) 033016,
  [\href{http://xxx.lanl.gov/abs/1306.6079}{{\tt 1306.6079}}].

\bibitem{PalomaresRuiz:2007eu}
S.~Palomares-Ruiz and S.~Pascoli, {\it {Testing MeV dark matter with neutrino
  detectors}},  {\em Phys. Rev.} {\bf D77} (2008) 025025,
  [\href{http://xxx.lanl.gov/abs/0710.5420}{{\tt 0710.5420}}].

\bibitem{Ho:2012ug}
C.~M. Ho and R.~J. Scherrer, {\it {Limits on MeV Dark Matter from the Effective
  Number of Neutrinos}},  {\em Phys. Rev.} {\bf D87} (2013), no.~2 023505,
  [\href{http://xxx.lanl.gov/abs/1208.4347}{{\tt 1208.4347}}].

\bibitem{Boddy:2015efa}
K.~K. Boddy and J.~Kumar, {\it {Indirect Detection of Dark Matter Using
  MeV-Range Gamma-Ray Telescopes}},  {\em Phys. Rev.} {\bf D92} (2015), no.~2
  023533, [\href{http://xxx.lanl.gov/abs/1504.04024}{{\tt 1504.04024}}].

\bibitem{Gonzalez-Morales:2017jkx}
A.~X. Gonzalez-Morales, S.~Profumo, and J.~Reynoso-C\'ordova, {\it {Prospects
  for indirect MeV Dark Matter detection with Gamma Rays in light of Cosmic
  Microwave Background Constraints}},  {\em Phys. Rev.} {\bf D96} (2017), no.~6
  063520, [\href{http://xxx.lanl.gov/abs/1705.00777}{{\tt 1705.00777}}].

\bibitem{Chen:2016tdz}
C.-S. Chen, G.-L. Lin, Y.-H. Lin, and F.~Xu, {\it {The 17 MeV Anomaly in
  Beryllium Decays and $U(1)$ Portal to Dark Matter}},  {\em Int. J. Mod.
  Phys.} {\bf A32} (2017), no.~31 1750178,
  [\href{http://xxx.lanl.gov/abs/1609.07198}{{\tt 1609.07198}}].

\bibitem{Fayet:2006sa}
P.~Fayet, D.~Hooper, and G.~Sigl, {\it {Constraints on light dark matter from
  core-collapse supernovae}},  {\em Phys. Rev. Lett.} {\bf 96} (2006) 211302,
  [\href{http://xxx.lanl.gov/abs/hep-ph/0602169}{{\tt hep-ph/0602169}}].

\bibitem{PhysRevD.73.103518}
Y.~Rasera, R.~Teyssier, P.~Sizun, M.~Cass\'e, P.~Fayet, B.~Cordier, and
  J.~Paul, {\it Soft gamma-ray background and light dark matter annihilation},
  {\em Phys. Rev. D} {\bf 73} (May, 2006) 103518.

\bibitem{Dreiner:2013mua}
H.~K. Dreiner, J.-F. Fortin, C.~Hanhart, and L.~Ubaldi, {\it {Supernova
  constraints on MeV dark sectors from $e^+e^-$ annihilations}},  {\em Phys.
  Rev.} {\bf D89} (2014), no.~10 105015,
  [\href{http://xxx.lanl.gov/abs/1310.3826}{{\tt 1310.3826}}].

\bibitem{Hooper:2007tu}
D.~Hooper, M.~Kaplinghat, L.~E. Strigari, and K.~M. Zurek, {\it {MeV Dark
  Matter and Small Scale Structure}},  {\em Phys. Rev.} {\bf D76} (2007)
  103515, [\href{http://xxx.lanl.gov/abs/0704.2558}{{\tt 0704.2558}}].

\bibitem{Boehm:2003ha}
C.~Boehm, P.~Fayet, and J.~Silk, {\it {Light and heavy dark matter particles}},
   {\em Phys. Rev.} {\bf D69} (2004) 101302,
  [\href{http://xxx.lanl.gov/abs/hep-ph/0311143}{{\tt hep-ph/0311143}}].

\bibitem{Frere:2006hp}
J.~M. Frere, F.~S. Ling, L.~Lopez~Honorez, E.~Nezri, Q.~Swillens, and
  G.~Vertongen, {\it {MeV right-handed neutrinos and dark matter}},  {\em Phys.
  Rev.} {\bf D75} (2007) 085017,
  [\href{http://xxx.lanl.gov/abs/hep-ph/0610240}{{\tt hep-ph/0610240}}].

\bibitem{Fayet:2006sp}
P.~Fayet, {\it {Constraints on Light Dark Matter and U bosons, from psi,
  Upsilon, K+, pi0, eta and eta-prime decays}},  {\em Phys. Rev.} {\bf D74}
  (2006) 054034, [\href{http://xxx.lanl.gov/abs/hep-ph/0607318}{{\tt
  hep-ph/0607318}}].

\bibitem{Fayet:2007ua}
P.~Fayet, {\it {U-boson production in e+ e- annihilations, psi and Upsilon
  decays, and Light Dark Matter}},  {\em Phys. Rev.} {\bf D75} (2007) 115017,
  [\href{http://xxx.lanl.gov/abs/hep-ph/0702176}{{\tt hep-ph/0702176}}].

\bibitem{Borodatchenkova:2005ct}
N.~Borodatchenkova, D.~Choudhury, and M.~Drees, {\it {Probing MeV dark matter
  at low-energy e+e- colliders}},  {\em Phys. Rev. Lett.} {\bf 96} (2006)
  141802, [\href{http://xxx.lanl.gov/abs/hep-ph/0510147}{{\tt
  hep-ph/0510147}}].

\bibitem{Serpico:2004nm}
P.~D. Serpico and G.~G. Raffelt, {\it {MeV-mass dark matter and primordial
  nucleosynthesis}},  {\em Phys. Rev.} {\bf D70} (2004) 043526,
  [\href{http://xxx.lanl.gov/abs/astro-ph/0403417}{{\tt astro-ph/0403417}}].

\bibitem{Ge:2017mcq}
S.-F. Ge and I.~M. Shoemaker, {\it {Constraining Photon Portal Dark Matter with
  Texono and Coherent Data}},  \href{http://xxx.lanl.gov/abs/1710.10889}{{\tt
  1710.10889}}.

\bibitem{Pospelov:2007mp}
M.~Pospelov, A.~Ritz, and M.~B. Voloshin, {\it {Secluded WIMP Dark Matter}},
  {\em Phys. Lett.} {\bf B662} (2008) 53--61,
  [\href{http://xxx.lanl.gov/abs/0711.4866}{{\tt 0711.4866}}].

\bibitem{Pospelov:2008zw}
M.~Pospelov, {\it {Secluded U(1) below the weak scale}},  {\em Phys. Rev.} {\bf
  D80} (2009) 095002, [\href{http://xxx.lanl.gov/abs/0811.1030}{{\tt
  0811.1030}}].

\bibitem{Alexander:2016aln}
J.~Alexander {\em et.~al.}, {\it {Dark Sectors 2016 Workshop: Community
  Report}},  2016.
\newblock \href{http://xxx.lanl.gov/abs/1608.08632}{{\tt 1608.08632}}.

\bibitem{Gondolo:1990dk}
P.~Gondolo and G.~Gelmini, {\it {Cosmic abundances of stable particles:
  Improved analysis}},  {\em Nucl. Phys.} {\bf B360} (1991) 145--179.

\bibitem{Maity:2018dgy}
D.~Maity and P.~Saha, {\it {Connecting CMB anisotropy and cold dark matter
  phenomenology via reheating}},
  \href{http://xxx.lanl.gov/abs/1801.03059}{{\tt 1801.03059}}.

\bibitem{Slatyer:2012yq}
T.~R. Slatyer, {\it {Energy Injection And Absorption In The Cosmic Dark Ages}},
   {\em Phys. Rev.} {\bf D87} (2013), no.~12 123513,
  [\href{http://xxx.lanl.gov/abs/1211.0283}{{\tt 1211.0283}}].

\bibitem{Slatyer:2015kla}
T.~R. Slatyer, {\it {Indirect Dark Matter Signatures in the Cosmic Dark Ages
  II. Ionization, Heating and Photon Production from Arbitrary Energy
  Injections}},  {\em Phys. Rev.} {\bf D93} (2016), no.~2 023521,
  [\href{http://xxx.lanl.gov/abs/1506.03812}{{\tt 1506.03812}}].

\bibitem{Slatyer:2015jla}
T.~R. Slatyer, {\it {Indirect dark matter signatures in the cosmic dark ages.
  I. Generalizing the bound on s-wave dark matter annihilation from Planck
  results}},  {\em Phys. Rev.} {\bf D93} (2016), no.~2 023527,
  [\href{http://xxx.lanl.gov/abs/1506.03811}{{\tt 1506.03811}}].

\bibitem{Fortin:2009rq}
J.-F. Fortin, J.~Shelton, S.~Thomas, and Y.~Zhao, {\it {Gamma Ray Spectra from
  Dark Matter Annihilation and Decay}},
  \href{http://xxx.lanl.gov/abs/0908.2258}{{\tt 0908.2258}}.

\bibitem{Cirelli:2010xx}
M.~Cirelli, G.~Corcella, A.~Hektor, G.~Hutsi, M.~Kadastik, P.~Panci, M.~Raidal,
  F.~Sala, and A.~Strumia, {\it {PPPC 4 DM ID: A Poor Particle Physicist
  Cookbook for Dark Matter Indirect Detection}},  {\em JCAP} {\bf 1103} (2011)
  051, [\href{http://xxx.lanl.gov/abs/1012.4515}{{\tt 1012.4515}}]. [Erratum:
  JCAP1210,E01(2012)].

\bibitem{Sjostrand:2014zea}
T.~Sj{\"o}strand, S.~Ask, J.~R. Christiansen, R.~Corke, N.~Desai, P.~Ilten,
  S.~Mrenna, S.~Prestel, C.~O. Rasmussen, and P.~Z. Skands, {\it {An
  Introduction to PYTHIA 8.2}},  {\em Comput. Phys. Commun.} {\bf 191} (2015)
  159--177, [\href{http://xxx.lanl.gov/abs/1410.3012}{{\tt 1410.3012}}].

\bibitem{Strong:1998ck}
A.~W. Strong, H.~Bloemen, R.~Diehl, W.~Hermsen, and V.~Schoenfelder, {\it
  {Comptel skymapping: A New approach using parallel computing}},  {\em
  Astrophys. Lett. Commun.} {\bf 39} (1999) 209,
  [\href{http://xxx.lanl.gov/abs/astro-ph/9811211}{{\tt astro-ph/9811211}}].

\bibitem{Knodlseder:2005yq}
J.~Knodlseder {\em et.~al.}, {\it {The All-sky distribution of 511 keV
  electron-positron annihilation emission}},  {\em Astron. Astrophys.} {\bf
  441} (2005) 513--532, [\href{http://xxx.lanl.gov/abs/astro-ph/0506026}{{\tt
  astro-ph/0506026}}].

\bibitem{Aharonian:2006wh}
{\bf H.E.S.S.} Collaboration, F.~Aharonian {\em et.~al.}, {\it {H.E.S.S.
  observations of the Galactic Center region and their possible dark matter
  interpretation}},  {\em Phys. Rev. Lett.} {\bf 97} (2006) 221102,
  [\href{http://xxx.lanl.gov/abs/astro-ph/0610509}{{\tt astro-ph/0610509}}].
  [Erratum: Phys. Rev. Lett.97,249901(2006)].

\bibitem{Abeysekara:2017jxs}
A.~U. Abeysekara {\em et.~al.}, {\it {A Search for Dark Matter in the Galactic
  Halo with HAWC}},  \href{http://xxx.lanl.gov/abs/1710.10288}{{\tt
  1710.10288}}.

\bibitem{Archambault:2017wyh}
{\bf VERITAS} Collaboration, S.~Archambault {\em et.~al.}, {\it {Dark Matter
  Constraints from a Joint Analysis of Dwarf Spheroidal Galaxy Observations
  with VERITAS}},  {\em Phys. Rev.} {\bf D95} (2017), no.~8 082001,
  [\href{http://xxx.lanl.gov/abs/1703.04937}{{\tt 1703.04937}}].

\bibitem{DeAngelis:2016slk}
{\bf e-ASTROGAM} Collaboration, A.~De~Angelis {\em et.~al.}, {\it {The
  e-ASTROGAM mission}},  {\em Exper. Astron.} {\bf 44} (2017), no.~1 25--82,
  [\href{http://xxx.lanl.gov/abs/1611.02232}{{\tt 1611.02232}}].

\bibitem{Boehm:2003bt}
C.~Boehm, D.~Hooper, J.~Silk, M.~Casse, and J.~Paul, {\it {MeV dark matter: Has
  it been detected?}},  {\em Phys. Rev. Lett.} {\bf 92} (2004) 101301,
  [\href{http://xxx.lanl.gov/abs/astro-ph/0309686}{{\tt astro-ph/0309686}}].

\bibitem{Hooper:2003sh}
D.~Hooper, F.~Ferrer, C.~Boehm, J.~Silk, J.~Paul, N.~W. Evans, and M.~Casse,
  {\it {Possible evidence for MeV dark matter in dwarf spheroidals}},  {\em
  Phys. Rev. Lett.} {\bf 93} (2004) 161302,
  [\href{http://xxx.lanl.gov/abs/astro-ph/0311150}{{\tt astro-ph/0311150}}].

\bibitem{Beacom:2004pe}
J.~F. Beacom, N.~F. Bell, and G.~Bertone, {\it {Gamma-ray constraint on
  Galactic positron production by MeV dark matter}},  {\em Phys. Rev. Lett.}
  {\bf 94} (2005) 171301.

\bibitem{Ahn:2005ck}
K.~Ahn and E.~Komatsu, {\it {Dark matter annihilation: The Origin of cosmic
  gamma-ray background at 1-20 -MeV}},  {\em Phys. Rev.} {\bf D72} (2005)
  061301, [\href{http://xxx.lanl.gov/abs/astro-ph/0506520}{{\tt
  astro-ph/0506520}}].

\bibitem{Lawson:2007kp}
K.~Lawson and A.~R. Zhitnitsky, {\it {Diffuse cosmic gamma-rays at 1-20 MeV: A
  trace of the dark matter?}},  {\em JCAP} {\bf 0801} (2008) 022,
  [\href{http://xxx.lanl.gov/abs/0704.3064}{{\tt 0704.3064}}].

\bibitem{Huh:2007zw}
J.-H. Huh, J.~E. Kim, J.-C. Park, and S.~C. Park, {\it {Galactic 511 keV line
  from MeV milli-charged dark matter}},  {\em Phys. Rev.} {\bf D77} (2008)
  123503, [\href{http://xxx.lanl.gov/abs/0711.3528}{{\tt 0711.3528}}].

\bibitem{Kahn:2007ru}
Y.~Kahn, M.~Schmitt, and T.~M.~P. Tait, {\it {Enhanced rare pion decays from a
  model of MeV dark matter}},  {\em Phys. Rev.} {\bf D78} (2008) 115002,
  [\href{http://xxx.lanl.gov/abs/0712.0007}{{\tt 0712.0007}}].

\bibitem{Queiroz:2014yna}
F.~S. Queiroz and K.~Sinha, {\it {The Poker Face of the Majoron Dark Matter
  Model: LUX to keV Line}},  {\em Phys. Lett.} {\bf B735} (2014) 69--74,
  [\href{http://xxx.lanl.gov/abs/1404.1400}{{\tt 1404.1400}}].

\bibitem{Mambrini:2015sia}
Y.~Mambrini, S.~Profumo, and F.~S. Queiroz, {\it {Dark Matter and Global
  Symmetries}},  {\em Phys. Lett.} {\bf B760} (2016) 807--815,
  [\href{http://xxx.lanl.gov/abs/1508.06635}{{\tt 1508.06635}}].

\bibitem{Bringmann:2016axu}
T.~Bringmann, A.~Galea, A.~Hryczuk, and C.~Weniger, {\it {Novel Spectral
  Features in MeV Gamma Rays from Dark Matter}},  {\em Phys. Rev.} {\bf D95}
  (2017), no.~4 043002, [\href{http://xxx.lanl.gov/abs/1610.04613}{{\tt
  1610.04613}}].

\bibitem{Boddy:2016fds}
K.~K. Boddy, K.~R. Dienes, D.~Kim, J.~Kumar, J.-C. Park, and B.~Thomas, {\it
  {Lines and Boxes: Unmasking Dynamical Dark Matter through Correlations in the
  MeV Gamma-Ray Spectrum}},  {\em Phys. Rev.} {\bf D94} (2016), no.~9 095027,
  [\href{http://xxx.lanl.gov/abs/1606.07440}{{\tt 1606.07440}}].

\bibitem{Garcia-Cely:2016pse}
C.~Garcia-Cely and J.~Heeck, {\it {Indirect searches of dark matter via
  polynomial spectral features}},  {\em JCAP} {\bf 1608} (2016) 023,
  [\href{http://xxx.lanl.gov/abs/1605.08049}{{\tt 1605.08049}}].

\bibitem{Garcia-Cely:2016hsk}
C.~Garcia-Cely and A.~Rivera, {\it {General calculation of the cross section
  for dark matter annihilations into two photons}},  {\em JCAP} {\bf 1703}
  (2017), no.~03 054, [\href{http://xxx.lanl.gov/abs/1611.08029}{{\tt
  1611.08029}}].

\bibitem{Brdar:2017wgy}
V.~Brdar, J.~Kopp, J.~Liu, and X.-P. Wang, {\it {Return of the X-rays: A New
  Hope for Fermionic Dark Matter at the keV Scale}},
  \href{http://xxx.lanl.gov/abs/1710.02146}{{\tt 1710.02146}}.

\bibitem{Bartels:2017dpb}
R.~Bartels, D.~Gaggero, and C.~Weniger, {\it {Prospects for indirect dark
  matter searches with MeV photons}},  {\em JCAP} {\bf 1705} (2017), no.~05
  001, [\href{http://xxx.lanl.gov/abs/1703.02546}{{\tt 1703.02546}}].

\bibitem{Navarro:1995iw}
J.~F. Navarro, C.~S. Frenk, and S.~D.~M. White, {\it {The Structure of cold
  dark matter halos}},  {\em Astrophys. J.} {\bf 462} (1996) 563--575,
  [\href{http://xxx.lanl.gov/abs/astro-ph/9508025}{{\tt astro-ph/9508025}}].

\bibitem{Navarro:1996gj}
J.~F. Navarro, C.~S. Frenk, and S.~D.~M. White, {\it {A Universal density
  profile from hierarchical clustering}},  {\em Astrophys. J.} {\bf 490} (1997)
  493--508, [\href{http://xxx.lanl.gov/abs/astro-ph/9611107}{{\tt
  astro-ph/9611107}}].

\bibitem{Aharonian:1981spy}
F.~A. Aharonian and A.~M. Atoyan, {\it {Cosmic gamma-rays associated with
  annihilation of relativistic e$^+$ - e$^−$ pairs}},  {\em Phys. Lett.} {\bf
  99B} (1981) 301--304.

\bibitem{Aharonian:2000iz}
F.~A. Aharonian and A.~M. Atoyan, {\it {Broad-band diffuse gamma-ray emission
  of the galactic disk}},  {\em Astron. Astrophys.} {\bf 362} (2000) 937,
  [\href{http://xxx.lanl.gov/abs/astro-ph/0009009}{{\tt astro-ph/0009009}}].

\bibitem{Abe:2015eos}
{\bf XMASS} Collaboration, K.~Abe {\em et.~al.}, {\it {Direct dark matter
  search by annual modulation in XMASS-I}},  {\em Phys. Lett.} {\bf B759}
  (2016) 272--276, [\href{http://xxx.lanl.gov/abs/1511.04807}{{\tt
  1511.04807}}].

\bibitem{Angloher:2015ewa}
{\bf CRESST} Collaboration, G.~Angloher {\em et.~al.}, {\it {Results on light
  dark matter particles with a low-threshold CRESST-II detector}},  {\em Eur.
  Phys. J.} {\bf C76} (2016), no.~1 25,
  [\href{http://xxx.lanl.gov/abs/1509.01515}{{\tt 1509.01515}}].

\bibitem{Agnese:2015nto}
{\bf SuperCDMS} Collaboration, R.~Agnese {\em et.~al.}, {\it {New Results from
  the Search for Low-Mass Weakly Interacting Massive Particles with the CDMS
  Low Ionization Threshold Experiment}},  {\em Phys. Rev. Lett.} {\bf 116}
  (2016), no.~7 071301, [\href{http://xxx.lanl.gov/abs/1509.02448}{{\tt
  1509.02448}}].

\bibitem{Aprile:2016swn}
{\bf XENON100} Collaboration, E.~Aprile {\em et.~al.}, {\it {XENON100 Dark
  Matter Results from a Combination of 477 Live Days}},  {\em Phys. Rev.} {\bf
  D94} (2016), no.~12 122001, [\href{http://xxx.lanl.gov/abs/1609.06154}{{\tt
  1609.06154}}].

\bibitem{Akerib:2016lao}
{\bf LUX} Collaboration, D.~S. Akerib {\em et.~al.}, {\it {Results on the
  Spin-Dependent Scattering of Weakly Interacting Massive Particles on Nucleons
  from the Run 3 Data of the LUX Experiment}},  {\em Phys. Rev. Lett.} {\bf
  116} (2016), no.~16 161302, [\href{http://xxx.lanl.gov/abs/1602.03489}{{\tt
  1602.03489}}].

\bibitem{Angloher:2016rji}
{\bf CRESST} Collaboration, G.~Angloher {\em et.~al.}, {\it {Dark-Photon Search
  using Data from CRESST-II Phase 2}},  {\em Eur. Phys. J.} {\bf C77} (2017),
  no.~5 299, [\href{http://xxx.lanl.gov/abs/1612.07662}{{\tt 1612.07662}}].

\bibitem{Fu:2016ega}
{\bf PandaX-II} Collaboration, C.~Fu {\em et.~al.}, {\it {Spin-Dependent
  Weakly-Interacting-Massive-Particle–Nucleon Cross Section Limits from First
  Data of PandaX-II Experiment}},  {\em Phys. Rev. Lett.} {\bf 118} (2017),
  no.~7 071301, [\href{http://xxx.lanl.gov/abs/1611.06553}{{\tt 1611.06553}}].

\bibitem{Akerib:2016vxi}
{\bf LUX} Collaboration, D.~S. Akerib {\em et.~al.}, {\it {Results from a
  search for dark matter in the complete LUX exposure}},  {\em Phys. Rev.
  Lett.} {\bf 118} (2017), no.~2 021303,
  [\href{http://xxx.lanl.gov/abs/1608.07648}{{\tt 1608.07648}}].

\bibitem{Amole:2017dex}
{\bf PICO} Collaboration, C.~Amole {\em et.~al.}, {\it {Dark Matter Search
  Results from the PICO-60 C$_3$F$_8$ Bubble Chamber}},  {\em Phys. Rev. Lett.}
  {\bf 118} (2017), no.~25 251301,
  [\href{http://xxx.lanl.gov/abs/1702.07666}{{\tt 1702.07666}}].

\bibitem{Aprile:2017ngb}
{\bf XENON} Collaboration, E.~Aprile {\em et.~al.}, {\it {Search for WIMP
  Inelastic Scattering off Xenon Nuclei with XENON100}},  {\em Phys. Rev.} {\bf
  D96} (2017), no.~2 022008, [\href{http://xxx.lanl.gov/abs/1705.05830}{{\tt
  1705.05830}}].

\bibitem{Witte:2017qsy}
S.~J. Witte and G.~B. Gelmini, {\it {Updated Constraints on the Dark Matter
  Interpretation of CDMS-II-Si Data}},  {\em JCAP} {\bf 1705} (2017), no.~05
  026, [\href{http://xxx.lanl.gov/abs/1703.06892}{{\tt 1703.06892}}].

\bibitem{Cui:2017nnn}
{\bf PandaX-II} Collaboration, X.~Cui {\em et.~al.}, {\it {Dark Matter Results
  From 54-Ton-Day Exposure of PandaX-II Experiment}},  {\em Phys. Rev. Lett.}
  {\bf 119} (2017), no.~18 181302,
  [\href{http://xxx.lanl.gov/abs/1708.06917}{{\tt 1708.06917}}].

\bibitem{Aprile:2017aty}
{\bf XENON} Collaboration, E.~Aprile {\em et.~al.}, {\it {The XENON1T Dark
  Matter Experiment}},  {\em Eur. Phys. J.} {\bf C77} (2017), no.~12 881,
  [\href{http://xxx.lanl.gov/abs/1708.07051}{{\tt 1708.07051}}].

\bibitem{Aprile:2017iyp}
{\bf XENON} Collaboration, E.~Aprile {\em et.~al.}, {\it {First Dark Matter
  Search Results from the XENON1T Experiment}},  {\em Phys. Rev. Lett.} {\bf
  119} (2017), no.~18 181301, [\href{http://xxx.lanl.gov/abs/1705.06655}{{\tt
  1705.06655}}].

\bibitem{Aprile:2017aas}
{\bf XENON} Collaboration, E.~Aprile {\em et.~al.}, {\it {Effective field
  theory search for high-energy nuclear recoils using the XENON100 dark matter
  detector}},  {\em Phys. Rev.} {\bf D96} (2017), no.~4 042004,
  [\href{http://xxx.lanl.gov/abs/1705.02614}{{\tt 1705.02614}}].

\bibitem{Aprile:2017kek}
{\bf XENON} Collaboration, E.~Aprile {\em et.~al.}, {\it {Search for magnetic
  inelastic dark matter with XENON100}},  {\em JCAP} {\bf 1710} (2017), no.~10
  039, [\href{http://xxx.lanl.gov/abs/1704.05804}{{\tt 1704.05804}}].

\bibitem{Aprile:2017yea}
{\bf XENON} Collaboration, E.~Aprile {\em et.~al.}, {\it {Search for Electronic
  Recoil Event Rate Modulation with 4 Years of XENON100 Data}},  {\em Phys.
  Rev. Lett.} {\bf 118} (2017), no.~10 101101,
  [\href{http://xxx.lanl.gov/abs/1701.00769}{{\tt 1701.00769}}].

\bibitem{Essig:2015cda}
R.~Essig, M.~Fernandez-Serra, J.~Mardon, A.~Soto, T.~Volansky, and T.-T. Yu,
  {\it {Direct Detection of sub-GeV Dark Matter with Semiconductor Targets}},
  {\em JHEP} {\bf 05} (2016) 046,
  [\href{http://xxx.lanl.gov/abs/1509.01598}{{\tt 1509.01598}}].

\bibitem{Essig:2017kqs}
R.~Essig, T.~Volansky, and T.-T. Yu, {\it {New Constraints and Prospects for
  sub-GeV Dark Matter Scattering off Electrons in Xenon}},  {\em Phys. Rev.}
  {\bf D96} (2017), no.~4 043017,
  [\href{http://xxx.lanl.gov/abs/1703.00910}{{\tt 1703.00910}}].

\bibitem{Scherrer:1985zt}
R.~J. Scherrer and M.~S. Turner, {\it {On the Relic, Cosmic Abundance of Stable
  Weakly Interacting Massive Particles}},  {\em Phys. Rev.} {\bf D33} (1986)
  1585. [Erratum: Phys. Rev.D34,3263(1986)].

\bibitem{Kamionkowski:1990ni}
M.~Kamionkowski and M.~S. Turner, {\it {THERMAL RELICS: DO WE KNOW THEIR
  ABUNDANCES?}},  {\em Phys. Rev.} {\bf D42} (1990) 3310--3320.

\bibitem{Hall:2009bx}
L.~J. Hall, K.~Jedamzik, J.~March-Russell, and S.~M. West, {\it {Freeze-In
  Production of FIMP Dark Matter}},  {\em JHEP} {\bf 1003} (2010) 080,
  [\href{http://xxx.lanl.gov/abs/0911.1120}{{\tt 0911.1120}}].

\bibitem{Blennow:2013jba}
M.~Blennow, E.~Fernandez-Martinez, and B.~Zaldivar, {\it {Freeze-in through
  portals}},  {\em JCAP} {\bf 1401} (2014) 003,
  [\href{http://xxx.lanl.gov/abs/1309.7348}{{\tt 1309.7348}}].

\bibitem{Klasen:2013ypa}
M.~Klasen and C.~E. Yaguna, {\it {Warm and cold fermionic dark matter via
  freeze-in}},  {\em JCAP} {\bf 1311} (2013) 039,
  [\href{http://xxx.lanl.gov/abs/1309.2777}{{\tt 1309.2777}}].

\bibitem{Shakya:2015xnx}
B.~Shakya, {\it {Sterile Neutrino Dark Matter from Freeze-In}},  {\em Mod.
  Phys. Lett.} {\bf A31} (2016), no.~06 1630005,
  [\href{http://xxx.lanl.gov/abs/1512.02751}{{\tt 1512.02751}}].

\bibitem{Bernal:2017kxu}
N.~Bernal, M.~Heikinheimo, T.~Tenkanen, K.~Tuominen, and V.~Vaskonen, {\it {The
  Dawn of FIMP Dark Matter: A Review of Models and Constraints}},  {\em Int. J.
  Mod. Phys.} {\bf A32} (2017), no.~27 1730023,
  [\href{http://xxx.lanl.gov/abs/1706.07442}{{\tt 1706.07442}}].

\bibitem{Davoudiasl:2014kua}
H.~Davoudiasl, H.-S. Lee, and W.~J. Marciano, {\it {Muon $g−2$, rare kaon
  decays, and parity violation from dark bosons}},  {\em Phys. Rev.} {\bf D89}
  (2014), no.~9 095006, [\href{http://xxx.lanl.gov/abs/1402.3620}{{\tt
  1402.3620}}].

\bibitem{Adler:2004hp}
{\bf E787} Collaboration, S.~Adler {\em et.~al.}, {\it {Further search for the
  decay K+ ---> pi+ nu anti-nu in the momentum region P < 195-MeV/c}},  {\em
  Phys. Rev.} {\bf D70} (2004) 037102,
  [\href{http://xxx.lanl.gov/abs/hep-ex/0403034}{{\tt hep-ex/0403034}}].

\bibitem{Artamonov:2008qb}
{\bf E949} Collaboration, A.~V. Artamonov {\em et.~al.}, {\it {New measurement
  of the $K^{+} \to \pi^{+} \nu \bar{\nu}$ branching ratio}},  {\em Phys. Rev.
  Lett.} {\bf 101} (2008) 191802,
  [\href{http://xxx.lanl.gov/abs/0808.2459}{{\tt 0808.2459}}].

\bibitem{Essig:2013vha}
R.~Essig, J.~Mardon, M.~Papucci, T.~Volansky, and Y.-M. Zhong, {\it
  {Constraining Light Dark Matter with Low-Energy $e^+e^-$ Colliders}},  {\em
  JHEP} {\bf 11} (2013) 167, [\href{http://xxx.lanl.gov/abs/1309.5084}{{\tt
  1309.5084}}].

\bibitem{Banerjee:2017hhz}
{\bf NA64} Collaboration, D.~Banerjee {\em et.~al.}, {\it {Search for vector
  mediator of Dark Matter production in invisible decay mode}},
  \href{http://xxx.lanl.gov/abs/1710.00971}{{\tt 1710.00971}}.

\bibitem{Battaglieri:2017aum}
M.~Battaglieri {\em et.~al.}, {\it {US Cosmic Visions: New Ideas in Dark Matter
  2017: Community Report}},  \href{http://xxx.lanl.gov/abs/1707.04591}{{\tt
  1707.04591}}.

\bibitem{Wojtsekhowski:2012zq}
B.~Wojtsekhowski, D.~Nikolenko, and I.~Rachek, {\it {Searching for a new force
  at VEPP-3}},  \href{http://xxx.lanl.gov/abs/1207.5089}{{\tt 1207.5089}}.

\bibitem{Raggi:2014zpa}
M.~Raggi and V.~Kozhuharov, {\it {Proposal to Search for a Dark Photon in
  Positron on Target Collisions at DA$\Phi$NE Linac}},  {\em Adv. High Energy
  Phys.} {\bf 2014} (2014) 959802,
  [\href{http://xxx.lanl.gov/abs/1403.3041}{{\tt 1403.3041}}].

\bibitem{Raggi:2015gza}
M.~Raggi, V.~Kozhuharov, and P.~Valente, {\it {The PADME experiment at LNF}},
  {\em EPJ Web Conf.} {\bf 96} (2015) 01025,
  [\href{http://xxx.lanl.gov/abs/1501.01867}{{\tt 1501.01867}}].

\bibitem{Balewski:2014pxa}
J.~Balewski {\em et.~al.}, {\it {The DarkLight Experiment: A Precision Search
  for New Physics at Low Energies}},  2014.
\newblock \href{http://xxx.lanl.gov/abs/1412.4717}{{\tt 1412.4717}}.

\bibitem{Wojtsekhowski:2017ijn}
B.~Wojtsekhowski {\em et.~al.}, {\it {Searching for a dark photon: Project of
  the experiment at VEPP-3}},  \href{http://xxx.lanl.gov/abs/1708.07901}{{\tt
  1708.07901}}.

\bibitem{Davoudiasl:2015vba}
H.~Davoudiasl, D.~Hooper, and S.~D. McDermott, {\it {Inflatable Dark Matter}},
  {\em Phys. Rev. Lett.} {\bf 116} (2016), no.~3 031303,
  [\href{http://xxx.lanl.gov/abs/1507.08660}{{\tt 1507.08660}}].

\bibitem{Hooper:2011aj}
D.~Hooper, F.~S. Queiroz, and N.~Y. Gnedin, {\it {Non-Thermal Dark Matter
  Mimicking An Additional Neutrino Species In The Early Universe}},  {\em Phys.
  Rev.} {\bf D85} (2012) 063513, [\href{http://xxx.lanl.gov/abs/1111.6599}{{\tt
  1111.6599}}].

\bibitem{DiBari:2013dna}
P.~Di~Bari, S.~F. King, and A.~Merle, {\it {Dark Radiation or Warm Dark Matter
  from long lived particle decays in the light of Planck}},  {\em Phys. Lett.}
  {\bf B724} (2013) 77--83, [\href{http://xxx.lanl.gov/abs/1303.6267}{{\tt
  1303.6267}}].

\bibitem{Kelso:2013paa}
C.~Kelso, S.~Profumo, and F.~S. Queiroz, {\it {Non-thermal WIMPs as "Dark
  Radiation" in Light of ATACAMA, SPT, WMAP9 and Planck}},  {\em Phys. Rev.}
  {\bf D88} (2013), no.~2 023511,
  [\href{http://xxx.lanl.gov/abs/1304.5243}{{\tt 1304.5243}}].

\bibitem{Kelso:2013nwa}
C.~Kelso, C.~A. de~S.~Pires, S.~Profumo, F.~S. Queiroz, and P.~S. Rodrigues~da
  Silva, {\it {A 331 WIMPy Dark Radiation Model}},  {\em Eur. Phys. J.} {\bf
  C74} (2014), no.~3 2797, [\href{http://xxx.lanl.gov/abs/1308.6630}{{\tt
  1308.6630}}].

\bibitem{Baer:2014eja}
H.~Baer, K.-Y. Choi, J.~E. Kim, and L.~Roszkowski, {\it {Dark matter production
  in the early Universe: beyond the thermal WIMP paradigm}},  {\em Phys. Rept.}
  {\bf 555} (2015) 1--60, [\href{http://xxx.lanl.gov/abs/1407.0017}{{\tt
  1407.0017}}].

\bibitem{Queiroz:2014ara}
F.~S. Queiroz, K.~Sinha, and W.~Wester, {\it {Rich tapestry: Supersymmetric
  axions, dark radiation, and inflationary reheating}},  {\em Phys. Rev.} {\bf
  D90} (2014), no.~11 115009, [\href{http://xxx.lanl.gov/abs/1407.4110}{{\tt
  1407.4110}}].

\bibitem{Allahverdi:2014bva}
R.~Allahverdi, B.~Dutta, F.~S. Queiroz, L.~E. Strigari, and M.-Y. Wang, {\it
  {Dark Matter from Late Invisible Decays to/of Gravitinos}},  {\em Phys. Rev.}
  {\bf D91} (2015), no.~5 055033,
  [\href{http://xxx.lanl.gov/abs/1412.4391}{{\tt 1412.4391}}].

\bibitem{Merle:2015oja}
A.~Merle and M.~Totzauer, {\it {keV Sterile Neutrino Dark Matter from Singlet
  Scalar Decays: Basic Concepts and Subtle Features}},  {\em JCAP} {\bf 1506}
  (2015) 011, [\href{http://xxx.lanl.gov/abs/1502.01011}{{\tt 1502.01011}}].

\bibitem{Aoki:2015nza}
M.~Aoki, T.~Toma, and A.~Vicente, {\it {Non-thermal Production of Minimal Dark
  Matter via Right-handed Neutrino Decay}},  {\em JCAP} {\bf 1509} (2015) 063,
  [\href{http://xxx.lanl.gov/abs/1507.01591}{{\tt 1507.01591}}].

\bibitem{Okada:2015kkj}
H.~Okada, Y.~Orikasa, and T.~Toma, {\it {Nonthermal dark matter models and
  signals}},  {\em Phys. Rev.} {\bf D93} (2016), no.~5 055007,
  [\href{http://xxx.lanl.gov/abs/1511.01018}{{\tt 1511.01018}}].

\bibitem{Kane:2015jia}
G.~Kane, K.~Sinha, and S.~Watson, {\it {Cosmological Moduli and the
  Post-Inflationary Universe: A Critical Review}},  {\em Int. J. Mod. Phys.}
  {\bf D24} (2015), no.~08 1530022,
  [\href{http://xxx.lanl.gov/abs/1502.07746}{{\tt 1502.07746}}].

\bibitem{Kim:2016spf}
H.~Kim, J.-P. Hong, and C.~S. Shin, {\it {A map of the non-thermal WIMP}},
  {\em Phys. Lett.} {\bf B768} (2017) 292--298,
  [\href{http://xxx.lanl.gov/abs/1611.02287}{{\tt 1611.02287}}].

\bibitem{Aparicio:2016qqb}
L.~Aparicio, M.~Cicoli, B.~Dutta, F.~Muia, and F.~Quevedo, {\it {Light Higgsino
  Dark Matter from Non-thermal Cosmology}},  {\em JHEP} {\bf 11} (2016) 038,
  [\href{http://xxx.lanl.gov/abs/1607.00004}{{\tt 1607.00004}}].

\bibitem{DEramo:2017gpl}
F.~D'Eramo, N.~Fernandez, and S.~Profumo, {\it {When the Universe Expands Too
  Fast: Relentless Dark Matter}},  {\em JCAP} {\bf 1705} (2017), no.~05 012,
  [\href{http://xxx.lanl.gov/abs/1703.04793}{{\tt 1703.04793}}].

\bibitem{Bramante:2017obj}
J.~Bramante and J.~Unwin, {\it {Superheavy Thermal Dark Matter and Primordial
  Asymmetries}},  {\em JHEP} {\bf 02} (2017) 119,
  [\href{http://xxx.lanl.gov/abs/1701.05859}{{\tt 1701.05859}}].

\bibitem{Dimastrogiovanni:2017tvd}
E.~Dimastrogiovanni and L.~M. Krauss, {\it {$\Delta N_{\text{eff}}$ and entropy
  production from early-decaying gravitinos}},
  \href{http://xxx.lanl.gov/abs/1706.01495}{{\tt 1706.01495}}.

\bibitem{Allahverdi:2017sks}
R.~Allahverdi, J.~B. Dent, and J.~Osinski, {\it {Non-thermal Production of Dark
  Matter from Primordial Black Holes}},
  \href{http://xxx.lanl.gov/abs/1711.10511}{{\tt 1711.10511}}.

\bibitem{Baur:2017stq}
J.~Baur, N.~Palanque-Delabrouille, C.~Yeche, A.~Boyarsky, O.~Ruchayskiy,
  E.~Armengaud, and J.~Lesgourgues, {\it {Constraints from Ly-$\alpha$ forests
  on non-thermal dark matter including resonantly-produced sterile neutrinos}},
   {\em JCAP} {\bf 1712} (2017), no.~12 013,
  [\href{http://xxx.lanl.gov/abs/1706.03118}{{\tt 1706.03118}}].

\bibitem{Hoof:2017ibo}
S.~Hoof and J.~Jaeckel, {\it {QCD axions and axions like particles in a
  two-inflation scenario}},  {\em Phys. Rev.} {\bf D96} (2017), no.~11 115016,
  [\href{http://xxx.lanl.gov/abs/1709.01090}{{\tt 1709.01090}}].

\bibitem{Agnese:2016cpb}
{\bf SuperCDMS} Collaboration, R.~Agnese {\em et.~al.}, {\it {Projected
  Sensitivity of the SuperCDMS SNOLAB experiment}},  {\em Phys. Rev.} {\bf D95}
  (2017), no.~8 082002, [\href{http://xxx.lanl.gov/abs/1610.00006}{{\tt
  1610.00006}}].

\bibitem{Agnese:2017njq}
{\bf SuperCDMS} Collaboration, R.~Agnese {\em et.~al.}, {\it {Results from the
  Super Cryogenic Dark Matter Search (SuperCDMS) experiment at Soudan}},
  \href{http://xxx.lanl.gov/abs/1708.08869}{{\tt 1708.08869}}.

\bibitem{Turner:1987by}
M.~S. Turner, {\it {Axions from SN 1987a}},  {\em Phys. Rev. Lett.} {\bf 60}
  (1988) 1797.

\bibitem{Hirata:1987hu}
{\bf Kamiokande-II} Collaboration, K.~Hirata {\em et.~al.}, {\it {Observation
  of a Neutrino Burst from the Supernova SN 1987a}},  {\em Phys. Rev. Lett.}
  {\bf 58} (1987) 1490--1493. [,727(1987)].

\bibitem{Bionta:1987qt}
R.~M. Bionta {\em et.~al.}, {\it {Observation of a Neutrino Burst in
  Coincidence with Supernova SN 1987a in the Large Magellanic Cloud}},  {\em
  Phys. Rev. Lett.} {\bf 58} (1987) 1494.

\bibitem{Kazanas:2014mca}
D.~Kazanas, R.~N. Mohapatra, S.~Nussinov, V.~L. Teplitz, and Y.~Zhang, {\it
  {Supernova Bounds on the Dark Photon Using its Electromagnetic Decay}},  {\em
  Nucl. Phys.} {\bf B890} (2014) 17--29,
  [\href{http://xxx.lanl.gov/abs/1410.0221}{{\tt 1410.0221}}].

\bibitem{Mahoney:2017jqk}
C.~Mahoney, A.~K. Leibovich, and A.~R. Zentner, {\it {Updated Constraints on
  Self-Interacting Dark Matter from Supernova 1987A}},  {\em Phys. Rev.} {\bf
  D96} (2017), no.~4 043018, [\href{http://xxx.lanl.gov/abs/1706.08871}{{\tt
  1706.08871}}].

\bibitem{smolinsky_dark_2017}
J.~Smolinsky and P.~Tanedo, {\it Dark photons from captured inelastic dark
  matter annihilation: Charged particle signatures},
  \href{http://xxx.lanl.gov/abs/1701.03168}{{\tt 1701.03168}}.

\bibitem{Raffelt:1996wa}
G.~G. Raffelt, {\em {Stars as laboratories for fundamental physics}}.
\newblock 1996.

\bibitem{Fradette:2014sza}
A.~Fradette, M.~Pospelov, J.~Pradler, and A.~Ritz, {\it {Cosmological
  Constraints on Very Dark Photons}},  {\em Phys. Rev.} {\bf D90} (2014), no.~3
  035022, [\href{http://xxx.lanl.gov/abs/1407.0993}{{\tt 1407.0993}}].

\end{thebibliography}\endgroup

\end{document}